\documentclass[usenatbib]{mnras}
\usepackage{amsmath,pdflscape,graphicx}

\defcitealias{PCO15}{Paper~I}
\defcitealias{PaFa16}{Paper~II}

\newcommand{\mean}[1]{\overline{#1}}
\newcommand{\vari}{\mean{\sigma}^2}
\newcommand{\stde}{\mean{\sigma}}
\newcommand{\sic}{\mean{\sigma}_\text{c}}
\newcommand{\diaf}{\text{d}}

\newcommand{\App}{Appendix~}
\newcommand{\Sm}{Section~}
\newcommand{\Sms}{Sections~}
\newcommand{\So}{\S}
\newcommand{\eqt}{Eq.~}
\newcommand{\eqto}{eq.~}
\newcommand{\tblo}{table~}
\newcommand{\fgr}{Fig.~}
\newcommand{\fgrs}{Figs~}
\newcommand{\fgro}{fig.~}
\newcommand{\fgros}{figs~}

\newcommand{\hdust}{\textsc{Hdust}}
\newcommand{\Ha}{\text{H}\,\alpha}
\newcommand{\Hb}{\text{H}\,\beta}
\newcommand{\Bg}{\text{Br}\,\gamma}

\newcommand{\Porb}{P_\text{orb}}
\newcommand{\qr}{q_\text{r}}
\newcommand{\ass}{\alpha_\text{ss}} % alpha from ShSu73
\newcommand{\QU}{QU}
\newcommand{\uqp}{(Q,U)}

\author[D.~Panoglou et al.]{%
 Despina Panoglou$^1$\thanks{\tt panoglou@on.br},
 Marcelo Borges Fernandes$^1$, Dietrich Baade$^2$, \and Daniel M.~Faes$^3$,
 Thomas Rivinius$^4$, Alex C.~Carciofi$^3$, Atsuo T.~Okazaki$^5$\\\\
 $^1$Observat\'orio Nacional, Rua General Jos\'e Cristino 77, S\~ao Crist\'ov\~ao RJ 20921-400, Rio de Janeiro, Brazil\\
 $^2$European Organisation for Astronomical Research in the Southern Hemisphere, Karl Schwarzschild-Str.\ 2, 85748 Garching bei \\M\"unchen, Germany\\
 $^3$Instituto de Astronomia, Geof\'isica e Ci\^encias Atmosf\'ericas, Universidade de S\~ao Paulo, Rua do Mat\~ao 1226, SP 05508-900, Brazil\\
 $^4$European Organisation for Astronomical Research in the Southern Hemisphere, Casilla, Santiago 19001, Chile\\
 $^5$Faculty of Engineering, Hokkai-Gakunen University, Sapporo, Hokkaido 062-8605, Japan
 }
\title[Polarisation variations in decretion discs]{Modelling the periodical variations in multiband polarisation and photometry for discs of binary Be stars}

\begin{document}
\maketitle
\begin{abstract}
The tidal interaction of a Be star with a binary companion forms two spiral arms that cause orbital modulation of the Be disc structure. The aim of this work is to identify observables in which this modulation is apparent. The structure of a Be disc in a coplanar circular binary system is computed with a smoothed-particle hydrodynamics code, and a radiation transfer code calculates the spectral energy distribution.
Line depolarisation was confirmed, with polarisation profiles nearly reverse to emission-line profiles. The continuum flux maximizes for pole-on discs, but photometric variability maximizes for edge-on discs. The linear polarisation exhibits one or two maxima per orbital cycle. While polarisation variability in visible passbands is important only at low inclinations, infrared bands may demonstrate high orbital variability even at large inclinations.
More evident is the modulation in the polarisation angle (PA) for low inclinations. The latter can be used to track azimuthal asymmetries for pole-on discs, where the spectroscopic variability in the violet-to-red (V/R) emission-component ratio disappears. PA reversals coincide with phases where V/R=1, tracking lines of sight directed towards regions where the approaching and receding arms overlap.
Continuum flux and polarisation are mostly in phase for neighbouring wavelength regions. It is suggested that studies of non-symmetric discs distorted by tidal forces from a secondary star may be used to study disc variabilities of other origins.
\end{abstract}


\section{Introduction}
Be stars are B-type stars surrounded by a transient circumstellar disc, and they typically show intrinsic polarisation up to {1.5\%} \citep{Yudi01}. The polarimetric signature of a Be star is mainly attributed to its disc and is influenced by the disc's physical and geometric properties, given that polarisation depends on the direction of each individual photon and its potential scattering during collisions with the disc material.

\cite{Okaz91} and \cite*{PaSa92} suggested that a precessing one-armed density wave of a circumstellar disc causes variations in the V/R ratio of double-peaked emission lines.
\cite{McDa00} found a correlation between V/R ratio and B-band polarisation of two Be shell stars whose discs were considered as having one-armed spiral structures. \cite{HaJo13b} calculated the V-band polarisation level along one rotation of a spiral density enhancement that follows a simple logarithmic spiral function, and showed that linear polarisation changes with the phase of the density-wave period.

When the disc of a binary Be star is tidally distorted, it develops two (generally different in structure) spiral arms \citep[e.g.][hereafter \citetalias{PCO15}]{OkBa02,PCO15}. It has been shown that the V/R cycle of line emission in such stars may be locked to the orbital period \citep[hereafter cited as \citetalias{PaFa16}]{PaFa16}, a phenomenon that had already been implied by some observations.
\cite{Coyn70} demonstrated phase-locked variations of the visual light curve and the linear polarisation at six optical wavelengths for the binary star $\beta$ Lyr, which is surrounded by a conventional accretion disc. \cite{StOk07} showed that V/R cycles occur in Be stars dominated by one-armed density waves (i.e.\ $\zeta$~Tau; \citealt{SRCl09}), as well as in stars with discs presumably consisting of two-armed waves due to binarity (i.e.\ 4 Her; \citealt{StOk07}).
These facts incited the conjecture that the existence of spiral arms in a Be disc causes variability locked to the rotational phase of the spiral arms in polarimetric properties as that seen in other observables.

Both binary interaction and global disc oscillations produce azimuthal asymmetries in the disc, resulting in variability of observables with period equal to the time of one rotation of the one- or two-armed spiral structure. As in \citetalias{PaFa16}, we will demonstrate that binary Be discs are ideal laboratories for the study of asymmetries in circumstellar structures and the associated variability.
Be stars exhibit variability both monotonic and periodic, in various timescales and of different origins. By simulations of the periodic variability caused by the tidal interaction with a binary companion, it is possible to derive conclusions for other nearly cyclic variations caused by disc asymmetries in the form of helical disc structures.

The outcome of this study is presented as follows: In \Sm\ref{s:mod} we describe the procedure followed to quantify the polarisation of Be discs in \Sm\ref{s:res}. In \Sm\ref{s:dis} we discuss certain aspects of our findings, while in \Sm\ref{s:con} the main conclusions are given and future potentials are identified.
Throughout the text, our results are juxtaposed with observational and modelling studies of other authors. Notes on the qualification of the shape of $\QU$ diagrams are given in \App\ref{s:how}.

\section{Methodology} \label{s:mod}
In \citetalias{PCO15}, three-dimensional SPH (smoothed-particle hydrodynamics) calculations were conducted with the code of \cite{OkBa02} for coplanar circular and eccentric Be binary systems. The decretion disc surrounding the Be star is considered purely gaseous. From the computed disc structure, the truncation radius and inner-disc power-law drop-off exponent were given as functions of the azimuthal angle (see \citetalias{PCO15} for definitions).
The steady rotating disc structure was confirmed for circular binaries, portrayed as a shifting in the azimuthal direction by an amount proportional to the phase difference. In \citetalias{PaFa16} this property was used in order to calculate the line emission across the orbital cycle using the Monte Carlo radiative transfer code {\hdust} \citep{CaBj06}.

The results for various sets of orbital and disc parameters presented in \citetalias{PaFa16} showed that emission-line profiles in Be binary systems can be used to constrain the observational and binary characteristics, and even confirm binarity in suspected binaries. Along one orbital cycle there is one blueward and one redward shift of the emission-line profiles. When such transitions occur in discs seen under intermediate inclination angles, we may find flat-topped and triple-peaked profiles.
When the disc is seen edge-on, there are two maxima of the V/R ratio per cycle and the variability amplitude of the V/R ratio of $\Ha$ maximizes, a fact that is more obvious for low-viscosity (and denser, if all other parameters are kept constant) discs. Further potential expediency lies in the calculation of other line profile observables (e.g.\ equivalent width, peak separation) and polarimetric observational predictions.

In this paper we present the first calculations of polarimetric observables for discs in Be binaries. The method and geometry of the system are the same as in \citetalias{PaFa16} (\So2 and \fgro3, respectively, in that paper). Circular polarisation is omitted, because polarisation in Be discs is mainly due to Thomson scattering by free electrons \citep{Chan46a,CoKr69,ZeSe72}.
Thus the polarisation of light is considered linear, in which case its vectorial form can be unambiguously derived from the two Stokes parameters $Q$ and $U$ \citep{Sto852a,Chan60}.

\begin{figure}\centering
\includegraphics[clip,trim=0cm 123mm 98mm 2mm,scale=.71]{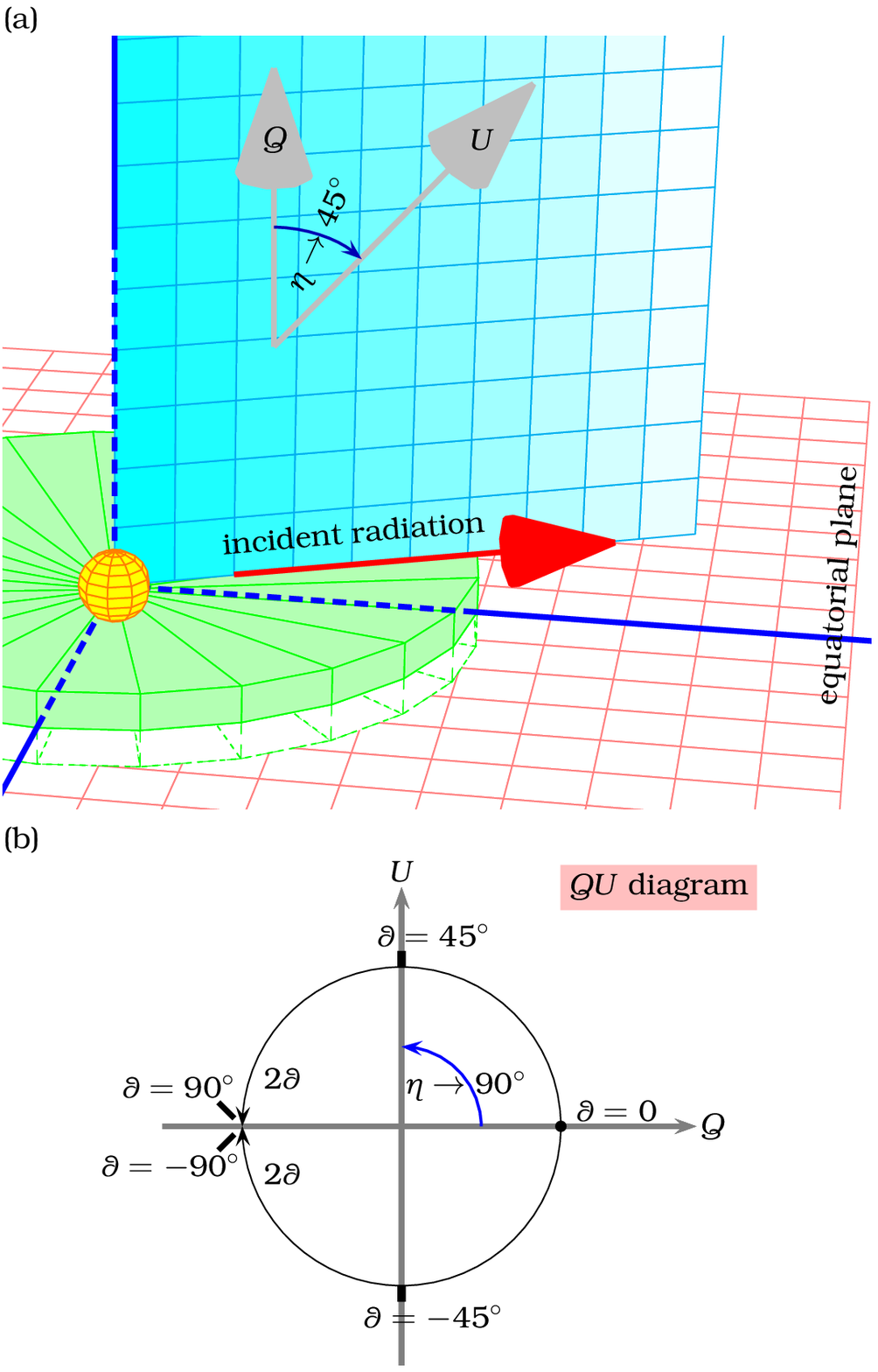}
\caption{(a) A stereographic representation of the system. The central star is denoted with a sphere surrounded by the disc on its equatorial plane. The $\QU$ plane contains the line of propagation of incident light. The $Q$ axis is vertical to the disc plane, and the angle $\eta$ between the $Q$ and $U$ axes is $45\degr$.\newline
(b)~Sample $\QU$ diagram. The $Q$ and $U$ axes form an orthogonal coordinate system, with $Q$ being the horizontal axis. Therefore the angle $\eta$ is here equal to $90\degr$, shown with an opposite direction with respect to its definition in the realistic sketch of the system geometry. The way to estimate the polarisation angle is overplotted, similar to that shown in \fgro3 of \protect\cite{MlCl79}.
The only difference is that angles in the range $[90,180]\degr$ are here denoted as being in the range $[-90,0]$\degr, in order to make it easier to clarify points of angle reversals. An angle $\theta\rightarrow90^-$ (approaching $90\degr$ from lower values) of \protect\citeauthor{MlCl79} becomes $\theta=90\degr$, while $\theta\rightarrow90^+$ (approaching $90\degr$ from larger values) is $\theta=-90\degr$ according to the hereby convention.}\label{f:geom}
\end{figure}

The linear polarisation is defined on a plane that passes from the line of propagation of incident radiation (\fgr\ref{f:geom}a; see also \citealt{BoHu83}). The normalised Stokes parameters comprise the total linear polarisation, whose amplitude $P$ is given by
  \begin{equation} P =\sqrt{Q^2+U^2}. \label{e:P} \end{equation}
Due to the transformation of the $\QU$ plane to its normalised equivalent (\fgr\ref{f:geom}b), the polarisation angle (PA; $\theta$) relative to the disc vertical is given by:
  \begin{equation}
  \text{PA} = \theta = \frac{1}{2}\arctan\frac{U}{Q}.
  \label{e:PA} \end{equation}
The Stokes components $Q$ and $U$, as well as the total polarisation $P$, are given as fractions of the totally emitted light (direct plus scattered starlight).

The polarised light that reaches the observer is a summation of the polarisation planes projected on a plane vertical to the line of sight. A spherically symmetric star with a spherical envelope would have zero polarisation, regardless of the direction from which it is observed. In principle, a disc would break the symmetry, except if it is perfectly axisymmetric and viewed pole-on.
In the latter case, the distribution of polarisation planes will also be uniform, and the polarisation of photons from orthogonal directions cancels out. In any other case, incomplete cancellations lead to non-zero net total polarisation. As the inclination angle increases, axisymmetric discs also show non-zero polarisation.
In general, polarisation arises at the occurrence of any apparent geometrical asymmetry, due to the observer's viewing angle and/or due to structural asymmetries of the star and its envelope. Both effects will be examined in the next sections.

\section{Results}\label{s:res}
The radiation transfer was computed for inclinations \mbox{$i\in[0,90]\degr$} spaced by 10\degr. Various spectral bands were tested, including all Johnson filters (\autoref{t:bands}). Results at some inclinations for a few of these bands will be presented, with the aim to draw an overall picture for the integrated flux, polarization level and PA, highlighting the main features.

\begin{table}\centering
\caption{Spectral bands referred to in the text and figures, with effective wavelength $\lambda_0$ and total width $w_0$. The second column denotes the respective label in figure legends.}\label{t:bands}
\begin{tabular}{ccccc}
\hline
  & label & $\lambda_0$ ($\micron$)    &$w_0$ ($\micron$) \\
\hline
  ultraviolet &             U &  0.357    &      0.109       & [1] \\
              &   Balmer jump &  0.364 \\
      visible &             B &  0.438    &      0.194       & [1] \\
      visible &             V &  0.547    &      0.251       & [1] \\
              &      H$\beta$ &  0.486    & $\lambda_0/50$  &      \\
              &     H$\alpha$ &  0.656    & $\lambda_0/50$  &      \\
      visible &             R &  0.670    &      0.417       & [1] \\
near-infrared &             I &  0.857    &      0.480       & [1] \\
              &  Paschen jump &  0.870 \\
near-infrared &             J &  1.210    &      0.581       & [1] \\
near-infrared &             K &  2.142    &      0.789       & [1] \\
near-infrared &           LII &  3.376    &      1.243       & [1] \\
\hline
\end{tabular}\\\par
[1] \url{http://svo2.cab.inta-csic.es/theory/fps/},\\ generic Johnson UBVRIJHKL system
\end{table}

We focus on low-viscosity discs, with the Shakura-Sunyaev turbulent viscosity parameter (simply called viscosity from here on; \citealt{ShSu73}) taken as \mbox{$\ass=0.1$}.
This is justified due to the dissipation time scale of a density wave that allows the formation of a non-axisymmetric disc \citep{PaLi95}, as well as from the simulations of \citetalias{PaFa16}, according to which high viscosities cause neither flat-topped and triple-peaked profiles nor significant V/R variability in azimuthally non-symmetric Be discs at any inclination.
Indeed, in high-viscosity discs line-emission profiles are symmetric and double-peaked (as in azimuthally symmetric discs), leaving no space for observable variability. A viscosity value equal or close to 0.1 was estimated for various Be stars in the past, e.g.\ \mbox{$\ass=0.1$} for $o$~And \citep*{ClTa03}, $\ass=0.14$ for 60~Cyg \citep{WiDr10}.

The stellar parameters used in the SPH simulation are: stellar mass \mbox{$M_*=14M_\odot$}, radius \mbox{$R_*=5R_\odot$}, effective temperature $T_\text{eff}=28000$~K (typical for a B0-1 star; \citealt{Harm88}). The binary parameters are: orbital period \mbox{$\Porb=94$~d} and secondary-to-primary mass ratio $\qr=0.14$.
This binary system is simulated with a constant disc feeding rate, resulting in a dense Be disc with density \mbox{$\rho_0=7\times10^{-11}$~g/cm$^3$} at the inner disc layer.
The parameters relevant to an oblate Be star are: gravity darkening parameter \mbox{$\beta=0.17$}, ratio of equatorial-to-polar radius $R_\text{eq}/R_\text{pol}=1.3$, and ratio of polar-to-equatorial temperature $T_\text{pol}/T_\text{eq}=1.3$. For more information on the system parameters, see \citetalias{PCO15} and \citetalias{PaFa16}.

\begin{figure}\centering
\includegraphics[clip,trim=3mm 3mm 2mm 2mm,scale=.9]{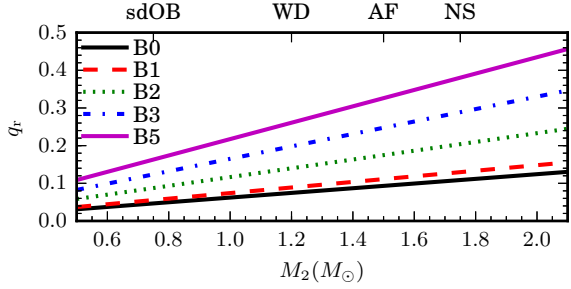}
\caption{The secondary-to-primary mass ratio ($\qr=M_2/M_1$) for typical Be binary systems, plotted against the mass $M_2$ of the secondary. Typical masses for the secondary star depending on its nature \citepalias{PCO15} are indicated on the top axis: sdO or sdB (sdOB), low-mass main-sequence A or F stars (AF), and compact objects, such as white dwarves (WD) and neutron stars (NS). High-mass B stars are considered as the primary component.
Values were taken mainly from \tblo3 of \citet{Harm88}.}\label{f:qr}
\end{figure}

The orbital parameters chosen are of the same order of magnitude as the parameters of most known Be binaries with varying nature of the secondary component. For instance, the binary B1\,V star $\pi$ Aqr (\mbox{$M_*/M_\odot=11\pm1.5$}, \mbox{$R_*/R_\odot=6.1\pm2.5$}) has an A or F main-sequence companion, $\Porb=84$ d and $\qr=0.16$ \citep{BjMi02}.
Be/X-ray binaries are usually of high total mass, mostly composed of a neutron star and a primary Be star of spectral type not later than B2 \citep{Reig11}. Their orbital period ranges from a few dozens to a few hundreds of days and the mass ratio is mostly between $[0.01,0.5]$ (\fgr\ref{f:qr}). The same range of values is also typical of Be binaries with a subdwarf O or B (sdO or sdB) companion \citep[see e.g.\ \tblo5 in][]{PePe13}.

The constant disc feeding rate is a prerequisite in order to have a steady disc in a circular binary, i.e.\ a disc that is neither dissipating nor building-up at the time scale of the orbital period. This way the disc is dynamically stable, i.e.\ there is always outward spiralling movement of matter, but the overall disc structure does not change.
Hence the orbitally-averaged quantities do not exhibit any monotonic changes over subsequent cycles, whilst the observed variations are simply caused by the changing viewing aspect of the binary system as it rotates.

In highly eccentric binaries, where the disc strength decreases close to the periastron passage and is maximal at apastron, the variations should also be periodical if the disc feeding rate remains constant \citepalias{PCO15}. However, the disc in that case is not dynamically stable any more and the observed variations will be the result of both the changing aspect and the varying disc structure.
Systematic modelling of long-period eccentric Be binaries could serve to study dissipating (from apastron to periastron) and growing discs (from the periastron towards apastron).

\subsection{Wavelength dependence}\label{s:wd}
Physical conditions (e.g.\ temperature) affect the ionization state and level populations of the circumstellar gas, so they can be tracked via the spectral energy distribution, line profiles and linear polarisation. Since the absorptive properties of matter change with wavelength, the dependence of polarisation upon wavelength may provide information on the density and chemical composition of the gas.

\begin{figure}\centering
\includegraphics[clip,trim=2mm 3mm 2mm 0mm,scale=.9]{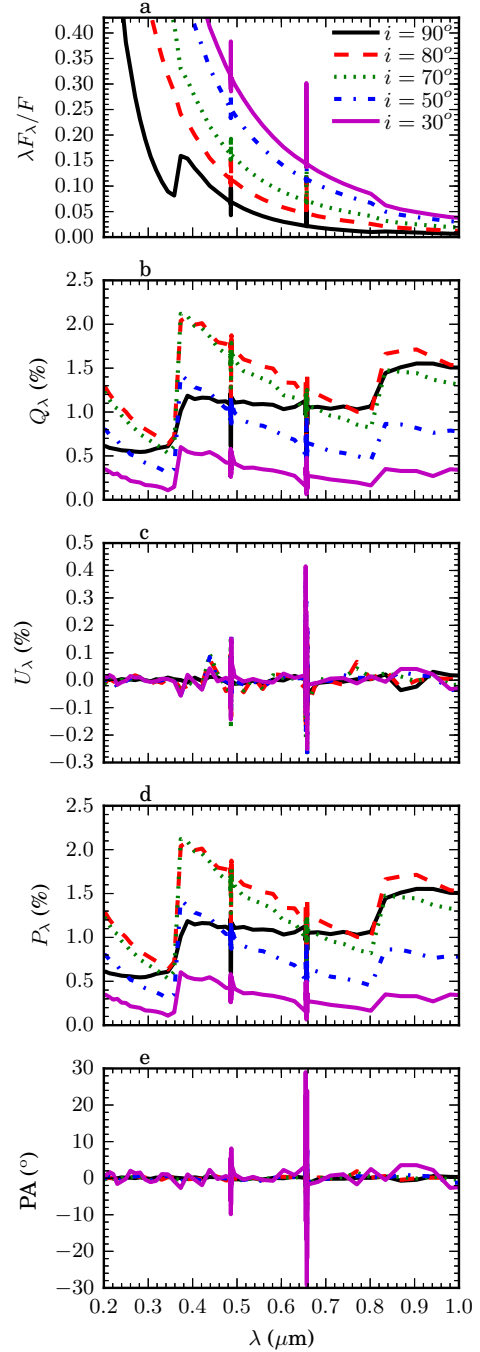}
\caption{From top to bottom: Continuum flux (a), the two orthogonal components of polarisation $Q$ and $U$ (b and c, respectively), linear polarisation $P$ (d), and PA (e), as functions of the wavelength, for five different inclination angles as labelled on the top panel. The Balmer and Paschen jumps are distinguishable at the designated wavelengths (\autoref{t:bands}), as well as the discontinuities at $\Ha$ and $\Hb$.}\label{f:fqu}
\end{figure}

The $Q$ axis is chosen perpendicular to the disc plane (i.e.\ parallel to the stellar rotation axis), and the $U$ axis is aligned at a $45\degr$ angle with respect to $Q$ (\fgr\ref{f:geom}a; \citealt{ViDr02}). The principal movement of the disc gas is to rotate around the star, hence the main change of polarisation occurs in the direction of disc rotation, i.e.\ vertical to the disc.
Therefore the total degree of polarisation is determined from $Q$, while $U$ has a mean value around zero for any inclination, as it traces stochastic movements of photons, Monte Carlo noise, and asymmetries caused by emission of recombination lines (see \fgrs\ref{f:fqu}b, c and d). This is directly connected to the fact that the PA is vertical to the Be disc plane \citep{QuBj97}.
So, when measured from the stellar rotational axis ($Q$), the PA is also around zero (\fgr\ref{f:fqu}e), and is expected to oscillate around zero in the course of one rotation, as will be shown in \Sm\ref{s:stat}. The PA does not change e.g.\ if the disk dissipates, whilst if the disc exhibits periodic variability, then the mean value of PA is zero. Therefore the PA already is an important quantity for the interpretation of other observables.

The sawtooth shape of polarisation spectra in \fgr\ref{f:fqu}(d) \citep[similar to \fgro1 of][]{HaJo13a} is due to abrupt changes (jumps) in the bound-free absorptive opacity of hydrogen at the limits of hydrogen series \citep*{PoBa79}. As a quantity, a ``jump'' originally measures the difference in continuum flux before and after the limiting wavelength \citep{Miha67}, as seen e.g.\ at \mbox{$i=90\degr$} in \fgr\ref{f:fqu}(a).
In this and other works that focus on polarisation, a jump refers to the difference in polarisation caused by the same effect.

Implementation of single scattering is adequate only in low-density environments, where the gas can be considered optically thin and net polarisation depends only on the total number of scatterers. In optically opaque discs, multiple scattering generally increases the polarisation level \citep[\fgros\mbox{8-9} of][]{WoBj96b}. At wavelengths longer than the Balmer jump (\mbox{$\sim3646$~\AA}), multiple scattering dominates and abruptly increases the polarisation.
At wavelengths shorter than the Balmer jump, the hydrogen opacity is large, most photons are absorbed instead of scattered, and the inclusion of multiple scattering does not modify the results with respect to pure single scattering by much. The same condition is evinced in \fgr\ref{f:fqu}(d) for the (smaller and less abrupt for the same inclination) Paschen jump at {$\sim8700$~\AA}, with the polarisation level decreasing for bands right before the jump and increasing after the jump.

\citet[their \fgro3]{HaJo13a} show that polarisation at {4000 \AA} (just after the Balmer jump) increases monotonically with the inclination angle, except for discs with large base densities $\rho_0>10^{-11}$ g/cm$^3$, where the maximum polarisation level lies at $i\simeq70\degr$. The figures in \cite{McDa00} also show a decrease in the computed B-band polarisation with increasing viewing angle in the range explored ($90\degr>i>80\degr$).
Maximum polarisation would normally be expected for edge-on discs ($i\rightarrow90\degr$), provided that they are optically thin and single scattering is dominant \citep{BrMl77}. At large optical depths however, scattered light is re-absorbed before it exits the disc, therefore multiple scattering is a more realistic approach \citep[\fgro8b]{WoBj96a}.
The effect of post-scattering attenuation is more pronounced for the light travelling parallel to the disc plane, hence polarisation is more intensely affected by multiple scattering for inclinations close to edge-on and denser discs. In general, polarisation increases for increasing $i$, but for a small range of inclinations close to $90\degr$ this effect is reversed, since the gas passes from a larger disc volume.
In our simulations (in which the disc lies at the high-density limit of \citealt{HaJo13a}), the maximum polarisation level at wavelength {$\lambda=4000$~\AA} appears at $i\simeq80\degr$ (\fgr\ref{f:fqu}).

\begin{figure}\centering
\includegraphics[clip,trim=2mm 3mm 0mm 1mm,scale=.9]{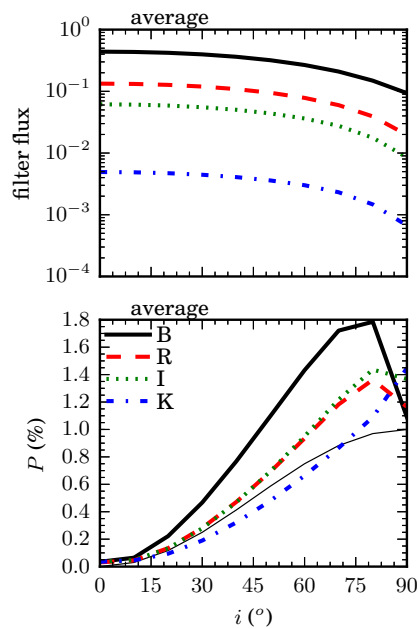}
\caption{The orbit-averaged values of photometric flux (top) and polarisation (bottom) as functions of the inclination angle $i$, for two visible (B, R) and two infrared (I, K) bands. The thin line on the bottom panel is the function $\sin^2i$, proportional to \eqto(1) of \protect\cite{Maga92}, which estimates the dependence of polarisation on $i$.}\label{f:smean}
\end{figure}

In \fgr\ref{f:smean} the mean values of flux and polarisation over a binary orbit are plotted as functions of $i$ for various visible and infrared filters. On the bottom panel of this figure, \eqto(1) in \citealt{Maga92} is also plotted, showing that it is a good estimation for the trend of polarisation with inclination \citep[see also][]{BrMl77}.
So polarisation is low at low inclinations, because the symmetry-breaking effect of the disc is smaller when viewed pole-on and increases towards edge-on orientations.
In their polarisation measurements of 48 Be stars (with a broad and a narrow filter around $\Ha$), \cite{PoMa76} also found little or no polarisation for pole-on stars. In our simulations, the mean value of linear polarisation maximizes at {80\degr} for all passbands, except for the infrared band K, in which polarisation steadily increases with inclination.

The simulations also show that at any wavelength the continuum flux is lower at higher inclinations. This is due to two effects of high inclinations: (a) the disc region that radiates towards the observer is smaller, and (b) more stellar flux is blocked by the disc.
Flux decreases from the ultraviolet to the infrared, and this might explain why photometric studies mostly involve the visible section of the electromagnetic spectrum (infrared passbands have fluxes at least an order of magnitude lower than visible bands). On the contrary, linear polarisation has comparable values in both visible and infrared passbands (i.e.\ not differring in each band by more than 0.5\% at any inclination).
Thus infrared polarimetry might be more valuable in studying circumstellar structures.

\subsubsection{Emission-line depolarisation}\label{s:elp}
Polarisation across recombination line profiles is sometimes assumed to be zero, but this can be a drawback in the interpretation of polarisation measurements, as discussed in \cite{MlCl79}. In cases of spherically non-symmetric structures, such as flattenned stars or circumstellar discs, velocity projections break the symmetry and zero polarisation is only an approximation.

\begin{figure}\centering
\includegraphics[clip,trim=2mm 3mm 0cm 0mm,scale=.9]{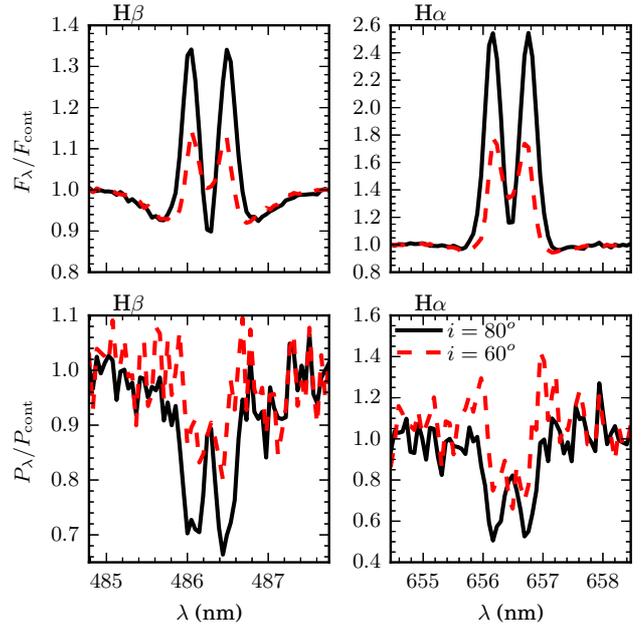}
\caption{Top: The radiation flux relative to continuum for $\Hb$ (left) and $\Ha$ (right) at inclination angles $i=60$ and $80\degr$ (labelled on the bottom right panel for both lines) at an arbitrary phase. Bottom: The linear polarisation relative to its continuum level for narrow bands around the same hydrogen lines.}\label{f:lha}
\end{figure}

In order to examine this phenomenon, we considered narrow passbands covering the full width (between \mbox{$\lambda_0\pm\lambda_0/100$}, where $\lambda_0$ is the effective wavelength; see \autoref{t:bands}) of $\Hb$ and $\Ha$, as opposed to broad-band filters. The resulting \fgr\ref{f:lha} shows that the computed linear polarisation decreases to 1/2 of the continuum polarisation in the wavelength range of $\Ha$ at $i=80\degr$.
\cite{PoBa79} also report that about half of their sample of Be stars show a decrease in polarisation to half of the continuum value across $\Ha$. A similar behaviour is seen in other hydrogen emission lines ($\Hb$, $\Bg$).

The phenomenon of recombination-line (de)polarisation is no novelty: \cite{ZeSe72} reported that narrow-band polarimetry of the shell star $\zeta$ Tau \citep{Bide76} shows that the $\Hb$ emission is ``considerably less polarized than the neighbouring continuum''. \cite{Coyn71} also noticed a minimum at polarisation of $\Ha$ in his polarisation survey of 28 Be stars.
In parallel, the increase of line polarisation relative to continuum (instead of de-polarisation) has been associated with shell absorption \citep{MlCl79}. More recently, high-resolution spectropolarimetric observations of the B[e] supergiant \citep{LaZi98} GG~Car revealed depolarisation at $\Ha$ from this line effect \citep{PeAr09}.

\cite{PoBa79} expected the ``depolarisation effect to be smaller'' in stars with lower average polarisation, which suggests that line polarisation might be better studied relative to continuum. If we defined violet and red peaks or equivalent widths (EW) for the (reverse) line polarisation profile, they all appear to have the same dependence as in the respective emission-line profile.
For instance, when the EW($F$) for the flux of $\Ha$ (or some other line) decreases, the EW($P$) of the corresponding polarisation profile decreases as well. Also, the central absorption of a double-peaked profile is known to deepen for shell stars ($i\rightarrow90\degr$), and in a similar manner the central core of a depolarisation profile increases at large inclinations. Similar trends are followed by the violet and red peak heights and their peak separation.

Just as the increased peak emission for double-peak profiles at large inclinations%
\footnote{The idea of emission increasing with $i$ might confuse readers familiar mainly with single-peak profiles. That is so because single-peak profiles become more pronounced at low inclinations, but for double-peak profiles (for stars with discs) the situation is different; see \citetalias{PaFa16} and \Sm\ref{s:lim}.}%
, the drop of the polarisation level $P$ at $\Ha$ becomes more intense for discs close to edge-on, while it decreases at low inclinations (\fgr\ref{f:lha}). \cite{BjNo91} found that \ion{Fe}{ii} lines in the ultraviolet and visible also exhibit depolarisation, and suggested that depolarisation is {``an attenuation effect arising from the removal of the polarised light} from the line of sight''.
In a demonstration of polarimetric diagnostics for circumstellar envelopes, \cite{NoBa92} confirm the latter effect, illustrated by the $\Ha$ line vanishing in a plot of flux multiplied by polarisation (a quantity that is often called ``polarised flux'').

It is thus certain that discontinuities of polarisation around recombination lines (\fgr\ref{f:fqu}d) are not caused by numerical noise due to the radiation transfer computation in those spectral regions, but it is a physical effect confirmed by observations. This was foreseen by \cite{PoBa79}, who correctly performed smoothing only on continuum polarisation data of the Be shell star EW Lac (their \fgro2), and not across the rapid changes of polarisation occurring at the Balmer lines.
Hence it might be worth to start studying line de/polarisation in the same detail that flux emission-line profiles are studied (i.e.\ define peak heights, full-width at half-maxima etc.). Of course this would be facilitated by high-resolution spectropolarimetric measurements.

Notably, the same number of photons was used for the computation of all curves in \fgr\ref{f:lha}, both for flux and polarisation fraction. It becomes obvious that radiation transfer codes need higher resolution (more photons for Monte Carlo codes, plus more bins per spectral range) to calculate polarisation than for flux, in order to reduce noise. Similarly, spectropolarimetric observations need higher spectral resolution in order to give well-defined polarisation profiles.
In any case, spectropolarimetry may be used complementarily to spectroscopy, e.g.\ for confirmation or in order to trigger spectroscopic monitoring. Alternatively, detected line emission might hint at depolarisation at the same wavelength.

The photons of the continuum light mainly come from the central source, while the photons of line emission originate in the circumstellar disc, i.e.\ a larger volume around the star. Hence the $\Ha$ photons scatter less often and thus are less polarised than the continuum, giving rise to line depolarisation \citep{ViDr02,ViHa05}.
However, there is no exact correspondence between flux profiles and profiles of the polarisation level $P$, since polarisation is a vector and carries more information than what is expressed by its amplitude ($P$) only. This is confirmed by many works which show that there is no unique transformation from the $\QU$ space to emission-line profiles in stars with circumstellar discs (e.g.\ \citealt{ViDr02} for Herbig Ae/Be stars; \citealt{HaKu09} for Herbig Ae/Be and classical Be stars).
While the theoretical study of $\QU$ diagrams across emission lines appears interesting, it is left for the future, as our current focus is on periodical variations of continuum polarisation.

\begin{figure*}\centering
\includegraphics[clip,trim=0mm 3mm 0cm 1mm,scale=.9]{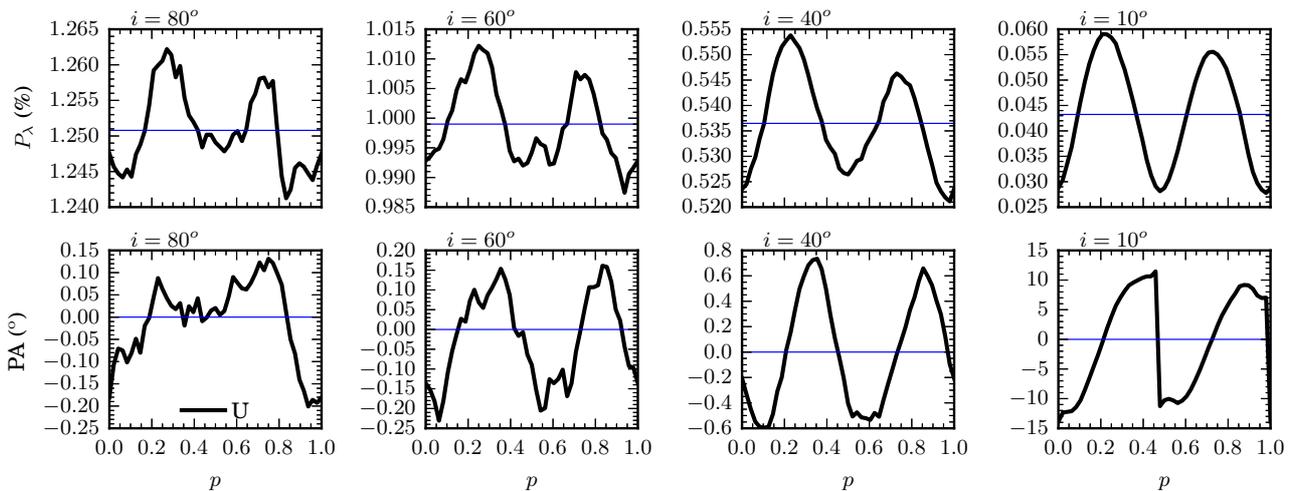}
\caption{Linear polarisation (top) and PA (bottom) over the U spectral range, as functions of the binary orbital phase $p$. The phase $p=0$ represents the position of the secondary star, so that each feature (e.g.\ the phase of maximum polarisation) appears with a respective phase difference with respect to the secondary's passing from the line of sight. On the upper panels the horizontal lines show the mean polarisation level, while on the bottom panels they mark the zero PA.}
\label{f:p}
\end{figure*}

\begin{figure*}\centering
\includegraphics[clip,trim=0mm 2mm 0cm 1mm,scale=.9]{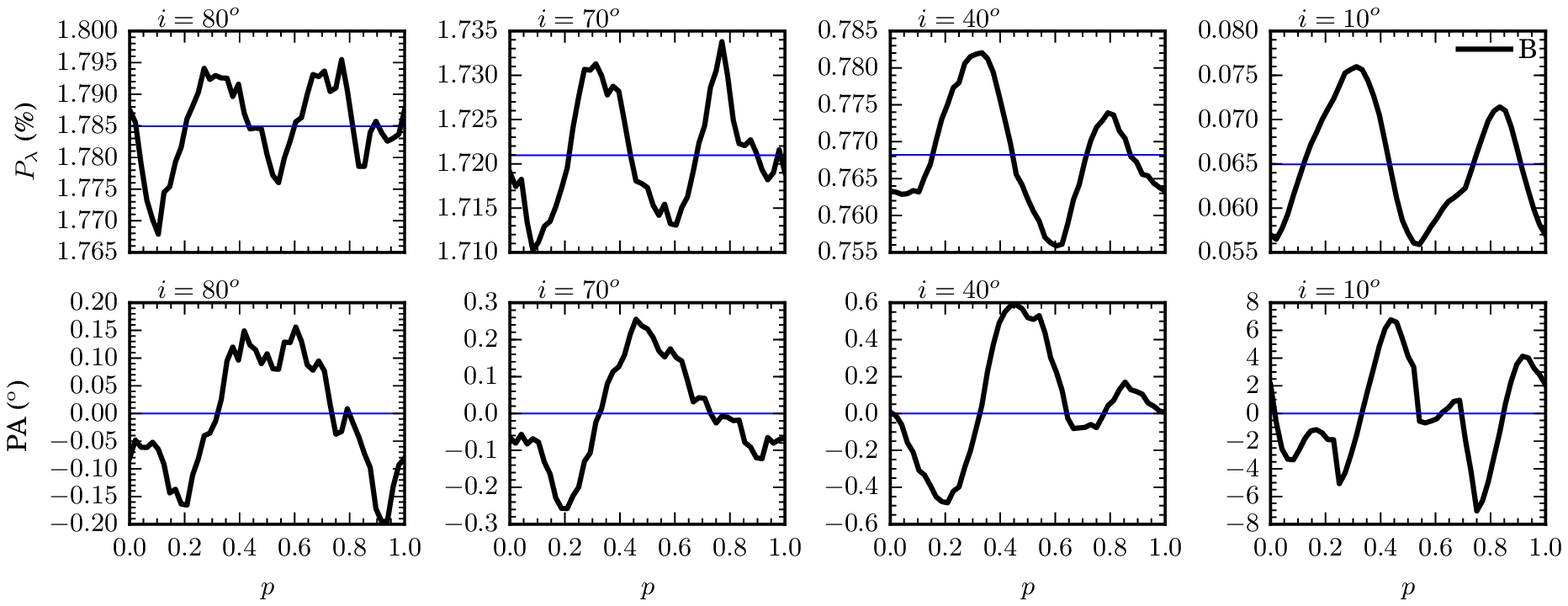}
\caption{Same as \fgr\ref{f:p}, but for the visible band B.}\label{f:p1}
\end{figure*}

\begin{figure*}\centering
\includegraphics[clip,trim=0mm 2mm 0cm 0mm,scale=.9]{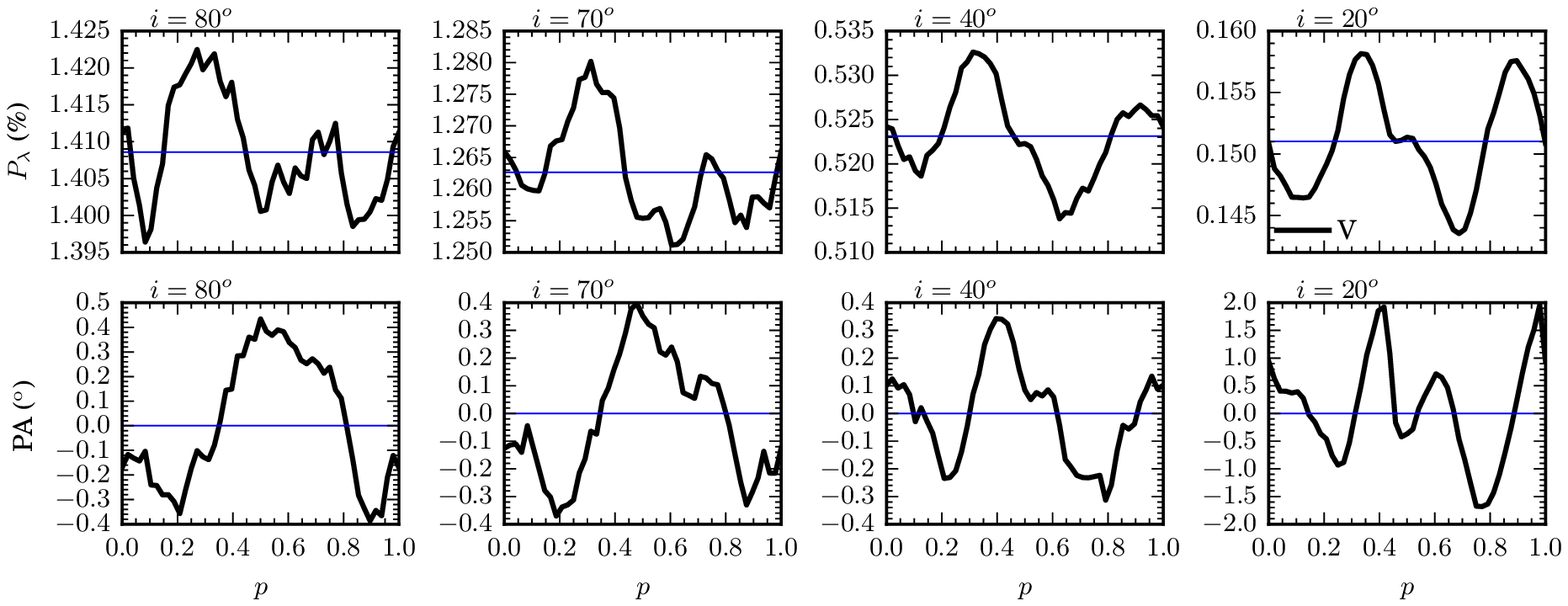}
\caption{Same as \fgr\ref{f:p}, but for the visible band V.}\label{f:pA}
\end{figure*}

\begin{figure*}\centering
\includegraphics[clip,trim=0mm 2mm 0cm 0mm,scale=.9]{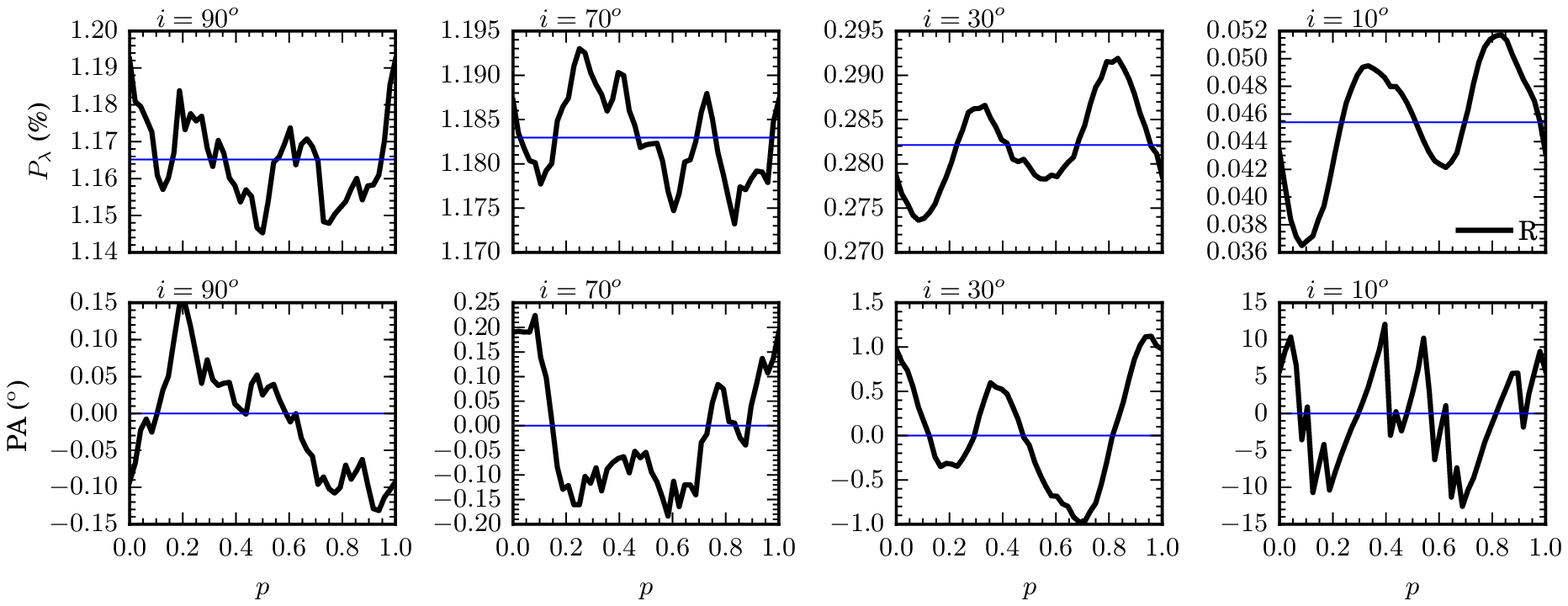}
\caption{Same as \fgr\ref{f:p}, but for the visible band R.}\label{f:pB}
\end{figure*}

\begin{figure*}\centering
\includegraphics[clip,trim=0mm 2mm 0cm 0mm,scale=.9]{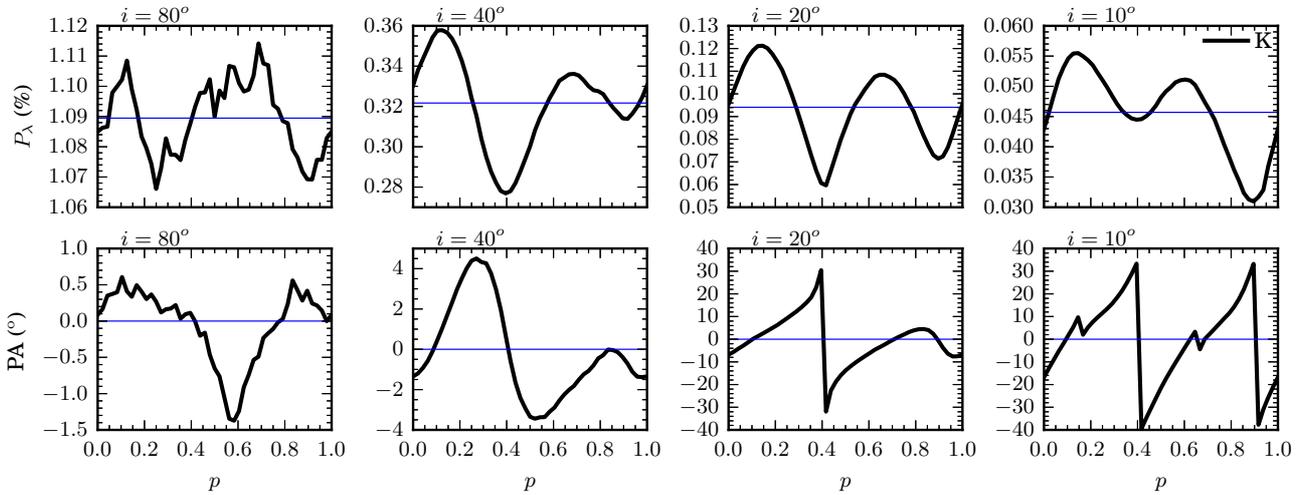}
\caption{Same as \fgr\ref{f:p}, but for the infrared band K.}\label{f:p22}
\end{figure*}

\subsection{Variability}\label{s:pd}
The polarimetric response with time is calculated at 50 different equidistant phases along one orbital cycle. At each phase the continuum quantities are integrated over the wavelength range.
\fgrs\ref{f:p}-\ref{f:p22} show the polarisation and PA (averaged over each band) as functions of the binary orbital phase for one ultraviolet (U), three visible (B, V, R) and one infrared (K) bands. For each filter, results at four inclinations are presented, selected such as to demonstrate the general trends.

The minimum, maximum and mean values of filter polarisation as functions of phase decrease from high to low disc inclinations. The PA acquires both positive and negative values, therefore the absolute values of PA extrema increase from high to low inclinations. Both $P$ and PA generally show one or two local maxima per period.
The orbital profile of PA may even show steep angle reversals (i.e.\ the PA abruptly flips on the opposite side of the disc) within the same cycle at orientations close to pole-on, especially in near-infrared bands. The PA is highly variable in the infrared for low inclinations, because near-infrared traces the inner disc, where azimuthal asymmetries are stronger \citep{WiKo07}.

At orientations close to pole-on the linear polarisation has two maxima of equal levels per period, while the PA reverses at the phases of the lowest polarisation. At higher inclinations, $P$ and PA tend to show one maximum (instead of two) per orbital period, as the amplitude of one of the maxima/minima decreases/increases.
This is due to one of the two spiral arms not being sufficiently dense to contribute to polarisation, since there is not much matter left to scatter light on the line of sight. Also, at high inclinations angle reversals tend to disappear, i.e.\ they become fewer per cycle and less steep.

\begin{figure}\centering
\includegraphics[clip,trim=2mm 3mm 2mm 1mm,scale=.9]{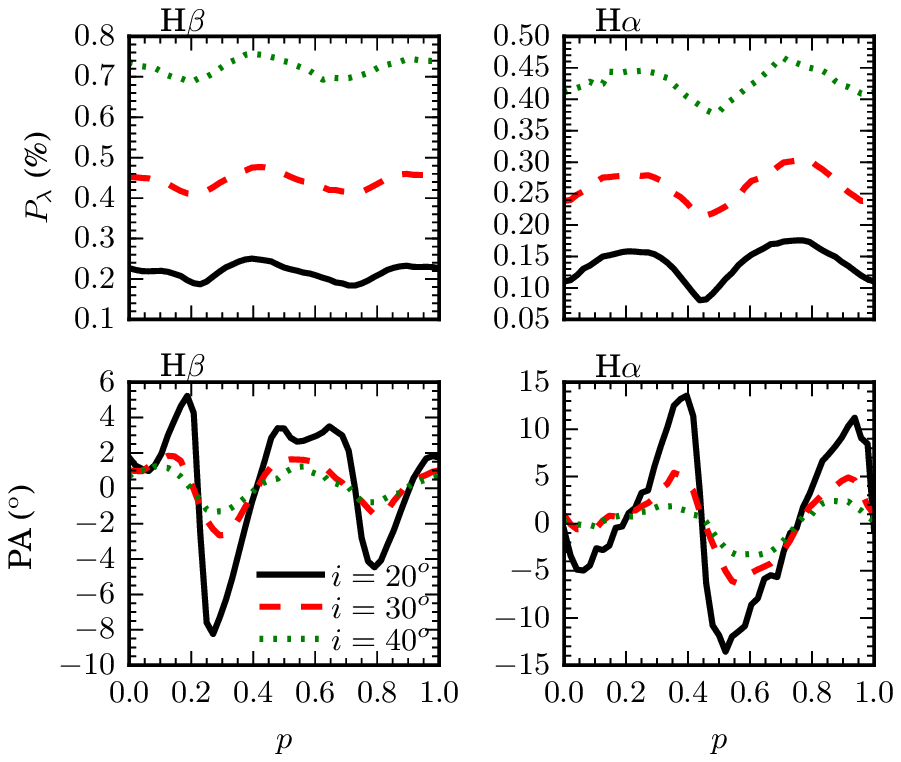}
\caption{Polarisation and PA averaged over narrow bands around $\Ha$ and $\Hb$, as functions of the binary orbital phase $p$ at three inclination angles (labelled on the bottom left panel).}\label{f:pl}
\end{figure}

\begin{figure*}\centering
\includegraphics[clip,trim=2mm 3mm 2mm 0mm,scale=.9]{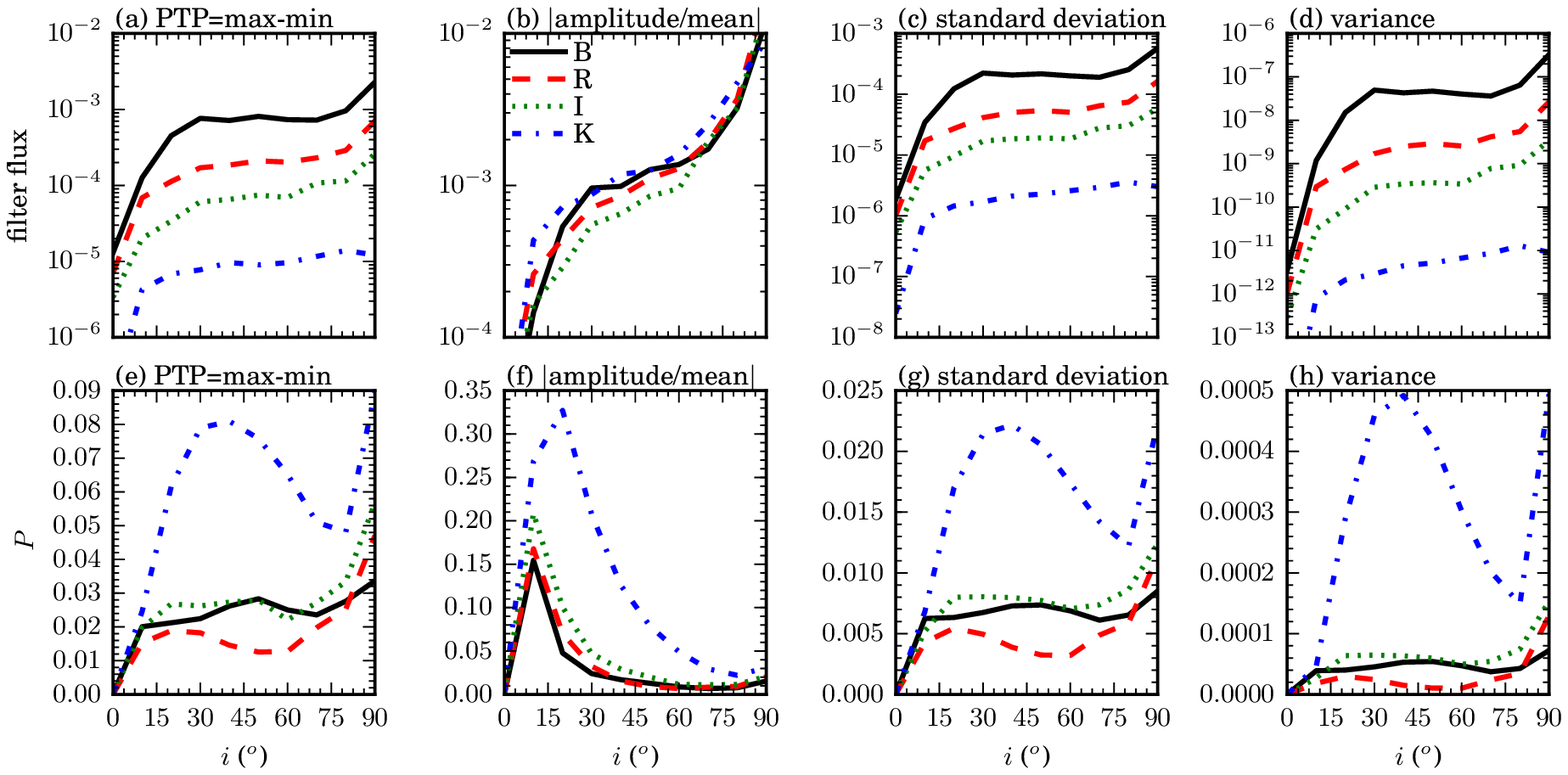}
\caption{The PTP difference over an orbital cycle (a, e), the ratio $R$ of amplitude to mean value (b, f), the mean standard deviation (c, g), and the mean variance $\vari$ (d, h) for the photometric flux (top) and linear polarisation (bottom), as functions of the inclination angle for the visible and infrared bands of \fgr\ref{f:smean}.}\label{f:sgen}
\end{figure*}

Observations cannot show angle reversals with certainty, unless we know the orientation of the disc with sufficient accuracy or if the observed difference is $\ge90\degr$ (such as the flip by $90\degr$ observed in V838 Mon by \citealt*{WiBj03}). However, given that the PA oscillates around zero both as a function of wavelength and as a function of time, the mean value of observed PA across wavelength can be considered as the zero-level.
The PA-versus-wavelength single-epoch panels for $\pi$~Aqr in \fgro9 of \cite{WiBj07} justify this assumption. Since $\pi$~Aqr has similar disc, stellar and orbital parameters as the system simulated in this work, it is possible to compare the mean polarisation computed here with the polarisation inferred over the UBVR range for $\pi$~Aqr by \citeauthor{WiBj07} ($0.5-2\%$).
\mbox{\fgrs\ref{f:smean} and \ref{f:p}-\ref{f:pB}} suggest that such polarisation levels correspond to an inclination of about 60\degr, as concluded also by \cite[$50\degr\le i\le75\degr$]{BjMi02}.

As noted in the previous section, emission-line depolarisation will probably be better studied as line emission flux, i.e.\ relative to continuum. It would then be more meaningful to explore variations in linear polarisation.
Nevertheless, two distinct observations of GG~Car (binary B[e] supergiant with orbital period 31 d with a dusty non-symmetric circumstellar disc; \citealt{HeLo81}, \citealt{Bofe10}, \citealt{MaBr12}) by \cite{PeAr09} show that the mean $P$ and PA around $\Ha$ both change in time and seem to be reversely correlated. \fgr\ref{f:pl} reproduces this behaviour for $\Ha$ and $\Hb$.
The description given previously for broad-band filters also applies here: the narrow-band polarisation level is higher for higher inclinations and shorter wavelengths, while the PA shows larger variability amplitudes for lower inclinations and longer wavelengths.

\subsubsection{Statistical analysis}\label{s:stat}
It is common to define the variability of a quantity $M$ in terms of amplitude. Amplitude is usually taken as the peak-to-peak (PTP) difference, i.e.\ the range of values, \mbox{$\text{PTP}(M) = \max M - \min M.\label{e:PTP}$} In this work, the amplitude of a variable quantity $M$ is defined as half of the PTP difference, $ A(M) = \text{PTP}(M)/2 $. The amplitude alone is an adequate measure of variability for quantities that oscillate around zero and take on both positive and negative values.
However, when there is an offset, i.e.\ when the quantity under study oscillates around a mean value $\mean{M}\neq0$, then the ratio $R$, defined as
  \begin{equation} R(M) = \frac{A(M)}{|\mean{M}|} \label{e:R}
  = \frac{\text{PTP}(M)}{2|\mean{M}|} ,\end{equation}
is a more appropriate measure of variability. Although both PTP and $R$ do not sample all points in the dataset, but only the minimum and maximum values, they are the only means to quantify variability when the available data is scarce.

The mean absolute difference of the quantity $M$ from its mean value $\mean{M}$ over one orbit is actually the mean standard deviation ($\stde$) of the variability, given by the square root of variance ($\vari$):
   \begin{equation} \stde(M) =\sqrt{\left<|M-\mean{M}|^2 \right>},
   \label{e:stde} \end{equation}
where the angular brackets denote averaged values. The variance of a variable is qualitatively similar to its PTP difference, but it is more accurate in characterising its variability, because it takes into account the differences of all distinct data points from their mean value.

\fgr\ref{f:sgen} shows the PTP difference, ratio $R$, standard deviation and variance for each filter's flux and $P$. Although the variability amplitude of the B-band flux shows a local maximum at $i\sim30\degr$, in fact it is not much compared to the mean value ($R\sim0.001$ or $10^{-5}\%$).
At large inclinations close to edge-on, the mean flux over an orbital cycle decreases for all passbands (\fgr\ref{f:smean}), but the PTP difference maximizes, so that the ratio $R$ steadily increases with increasing $i$ (\fgr\ref{f:sgen}b), in accord with line profile variability \citepalias{PaFa16}. However, even at $i=90\degr$, the variability in continuum flux, as expressed by the ratio $R$, does not exceed $0.01$ in any passband.
Such low photometric variability can be undetectable from the ground (except for very bright stars, if the signal-to-noise ratio is too high), but it might increase if azimuthal asymmetries within the disc become more violent, e.g.\ when larger density derivatives develop. Both the mean value and variability of continuum flux are maximal in the visible bands.

The PTP difference of linear polarisation generally increases with $i$, but its mean value also increases. It becomes clear that the PTP difference is not a good means to measure variability of periodic quantities. Variability in $P$ is more appropriately expressed by $R$, which shows a maximum at $i\in[10,30]\degr$. This particular behaviour was tested in various narrow and broad wavelength ranges and proved to be followed in all of them.
In visible bands, the PTP difference of linear polarisation is significant with respect to its mean value only for $i\in[10-20]\degr$, with the maximum PTP/mean ratio occurring at $i\sim10\degr$ ($R\sim15\%$). In the near-infrared band K, $R$ reaches a maximum (35\%) at $i=20\degr$, remains $>15\%$ up to $i=40\degr$, and is $\sim3\%$ even at 90\degr.
In general, the variability of $P$ in visible bands almost disappears at $i>30\degr$, while in infrared bands it is kept relatively high even at large inclinations.

\begin{figure}\centering
\includegraphics[clip,trim=2mm 3mm 2mm 1mm,scale=.9]{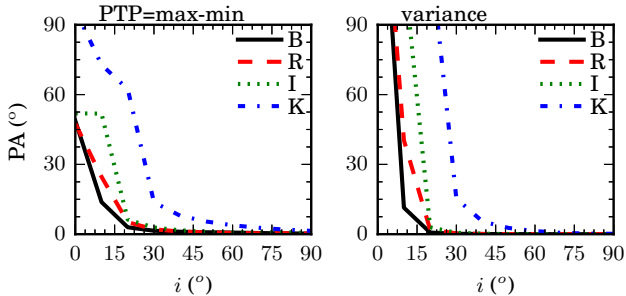}
\caption{The PTP difference (left) and the variance (right) of PA as functions of the inclination angle $i$. As the PTP difference for PA may reach up to 180\degr, while its variance may be well above even this high value (especially at low $i$), the range of the vertical axis in both panels was clipped to {[0,90]\degr} for clarity.}\label{f:varpa}
\end{figure}

\begin{figure*}\centering
\includegraphics[clip,trim=2mm 3mm 2mm 3mm,scale=.9]{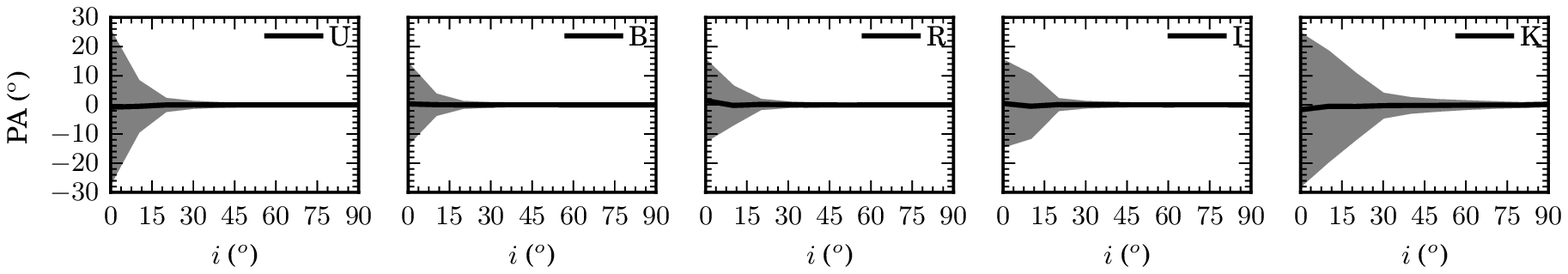}
\caption{The orbit-averaged value of PA (solid line) as a function of the inclination angle $i$ for various passbands. The shaded region shows the mean standard deviation over an orbital cycle at each inclination.}\label{f:stpa}
\end{figure*}

The polarisation angle is a quantity whose mean value is around zero, therefore the PTP difference alone is enough to express its variability (\fgr\ref{f:varpa}). The PTP difference for PA may reach up to {180\degr} in case of very steep angle reversals. Since the ratio $R$ is defined only for non-zero mean values, it has not been calculated for PA.
The variability in PA becomes more easy to qualify from \fgr\ref{f:stpa}, which shows the mean value of PA together with its mean standard deviation over an orbit. It becomes obvious that variability in PA is maximal at the lowest inclination (pole-on), and for most bands it disappears for $i>30\degr$. However, for the infrared-most bands some variability in PA exists also at large inclinations.

The polarimetric observations of \citet[UBVRI]{McDa99} for various Be stars support the above statistics, whenever the inclination is known. For instance, in agreement with the detailed description given above for low inclinations, the nearly pole-on star 48~Per \citep{QuBj97,DeSt11} has low but variable polarisation. The variability of 48 Per in PA is stronger than in $P$, but there seems to exist no preferential direction (i.e.\ mean value of PA is zero).
On the other hand, the edge-on star $\zeta$~Tau shows high variability in the polarisation level but nearly constant PA (as supported by the variance in $P$ and PA in \fgrs\ref{f:sgen} and \ref{f:varpa}).

Such considerations imply that it is possible to characterise and constrain variations in circumstellar discs of whatever origin, consulting simulations of non-axisymmetric discs formed by tidal interaction with a companion star. If the disc feeding rate is constant these variations should be periodical and locked to the orbital period of the binary system. Observations of long-term monotonic variations hint at changes in the disc strength.
Episodic changes that interrupt a monotonic behaviour are also possible, and they might cause confusion in the interpretation of observations, as discussed by \citet[\So4.1]{WiDr10}, especially with respect to the estimation of the outflow time scale.

\subsubsection{Phase differences}\label{s:fdi}
The phase lags between the maxima of $P$ and PA for different passbands resemble phase differences between the V/R ratios of various spectral lines.
\cite{WiKo07} showed that the timescale of V/R variations in $\zeta$ Tau is consistent with the timescale of rotation of an one-armed density wave \citep{Okaz97a}, while phase differences of the V/R ratio for $\Ha$ as compared to infrared lines were explained by the latter forming in a region of smaller radius than the region were $\Ha$ is excited (see e.g.\ \fgro3 of \citeauthor{WiKo07}).
Phase differences in the V/R ratio of different lines have been reproduced for a two-armed disc in \citetalias{PaFa16}, as a result (a) of the emitting region of each line having different sizes, and (b) of the azimuthal variation of disc density at the radial extent of the emitting region, due to the spiral arms receding from and approaching the line of sight. Similarly, polarisation phase lags in different bands are expected due to each one arising from a different disc region.

\begin{figure}\centering
\includegraphics[clip,trim=1mm 3mm 2mm 2mm,scale=.9]{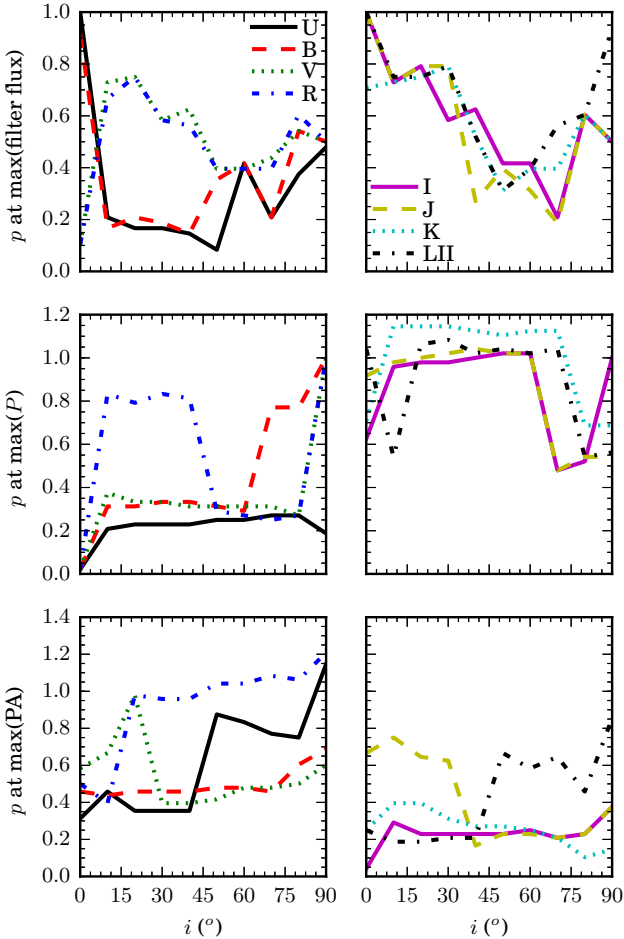}
\caption{The phase $p$ where the maximum value of any quantity $M$ (flux, $P$ and PA) occurs as a function of the inclination angle. Each column has plots for four different passbands labelled on the panels of the first row (ultraviolet/visible on the left, and near-infrared on the right).
Occasionally the phases were appropriately adjusted to an equivalent phase of an adjacent cycle ($p\leftarrow p+1$), so as to eliminate (a)~high derivatives of $\partial p(\max M)/\partial i$ and (b)~large differences between the $p$ at maxima for neighbouring bands. To compare phases at maxima of different panels, it might be needed to add or subtract $\Delta p=1$ from the phase $p$ that is plotted. The error in phase is \mbox{$|\Delta p|\simeq2/50=0.04$}.}
\label{f:pAm}
\end{figure}

The statistical analysis performed to evaluate mean values and amplitudes of the periodic quantities (\fgrs\ref{f:smean}, \ref{f:sgen} and \ref{f:varpa}) facilitated the estimation of the phase $p$ where the maximum and minimum values are met along an orbital cycle for each filter. \fgr\ref{f:pAm} plots the phase at maximum flux, $P$ and PA for all filters listed in \autoref{t:bands}. The maxima of filters close to the Balmer jump (U, B) almost coincide, especially for flux and polarisation.
A similar condition holds for near-infrared bands, showing maximum flux and $P$ at about the same phase for all inclinations. Computed phase differences of about $\Delta p=0.5$ indicate that the two maxima have similar values and the code catches the maximum corresponding to the other spiral arm.
Emission-line polarisation and PA for narrow spectral bands around $\Ha$ and $\Hb$ are characterised by rather large phase differences ($\sim\upi/2$ or $\Delta p>0.2$) along the orbital cycle (\fgr\ref{f:pl}), even though both lines belong to the visible section of the electromagnetic spectrum.

Another effect that can be explored is the phase differences between the maxima of flux, polarisation and PA at the same wavelength or passband. A closer look at \fgr\ref{f:pAm} reveals, for instance, that the phase differences at maximum flux and polarisation reaches up to $\Delta p=0.5$ at high inclinations in the visible. There are also lags between continuum and spectroscopic features, and they appear to increase for close to edge-on orientations.
Time lags of \mbox{1-3} months between polarisation and spectroscopic features have been observed for the Be stars $\gamma$~Cas \citep[binary system with \mbox{$\Porb=204~\text{d}\simeq7$} months;][]{PoMa78a} and $o$~And \citep{PoBa79}. Assuming that linear polarisation originates within three stellar radii from the central Be star, \cite{PoMa78a} estimate that the observed time lag for $o$~And agrees with the time needed for the gas to reach the formation region of shell lines.

Although the discs of $o$ And and $\gamma$ Cas are considered dissipating discs, the estimations of the above time scales are based on the assumption of viscous diffusion. A similar approach has been used by \cite{WiDr10} for the disc surrounding 60 Cyg, in order to estimate the value of viscosity from the time lag between the minima and maxima of $\text{EW}(\Ha)$.
This concept shows the relevance with this work, where the disc might not be dissipating but the gas indeed flows outwards due to viscous diffusion \citep*{LeOs91}. In both cases the viscous force induces the radial velocity component.

The results are quite complex and a detailed study of each quantity (flux, $P$, PA) for different passbands at any specified inclination should be performed in parallel to the exploration of the parameter space.
Although beyond the scope of this paper, the outcome of such a study will probably be able to elucidate many uncertainties about the light reaching the observer from discs distorted by binary interaction, and as a consequence from any disc whose structure varies both radially and azimuthally.

\subsubsection{$QU$ diagrams}\label{s:qU}
\begin{figure*}\centering
\includegraphics[clip,trim=4mm 8mm 0mm 0mm,scale=.9]{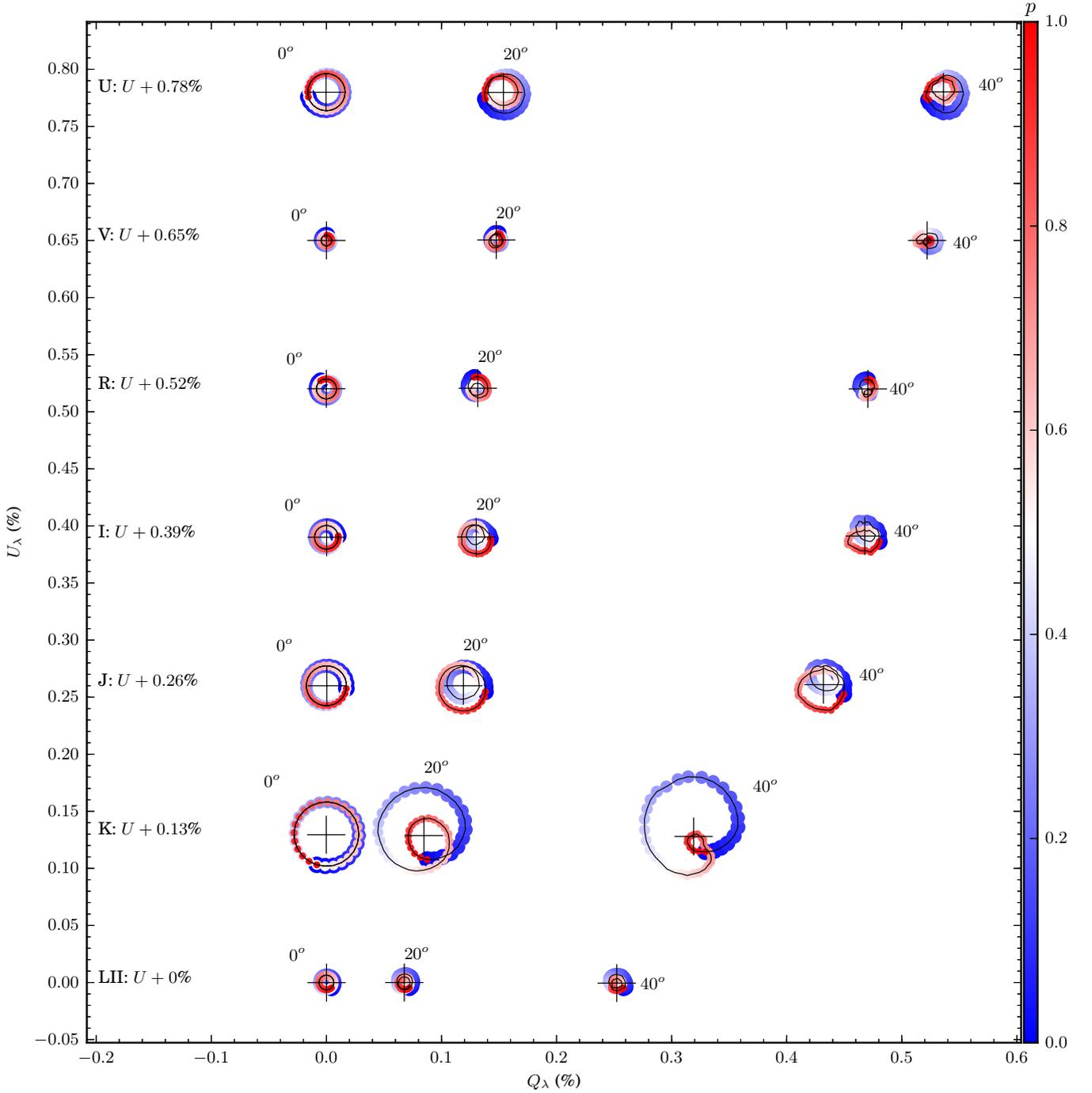}
\caption{Orbital cycles in the $QU$ diagram for $Q$ and $U$ averaged over the spectral band, at inclinations $i=0,20,40\degr$. Each (horizontal) series of $\QU$ curves corresponds to the results at different inclinations for the same band, with the $U$ polarisation component given with a positive offset, as labelled on the left. The observational inclination $i$ is denoted above each $\QU$ curve.
The cross sign superimposed on each curve marks the central point $\text{C}\equiv(\mean{Q},\mean{U})$ of the cycle (see text). Each shaded circle in a $\QU$ curve corresponds to one phase $p$ along the orbital cycle. The size of each circle decreases for increasing $p$, while its shading intensity also changes: fainter circles take place around $p=0.5$. The plot axes have the same aspect ratio, i.e.\ they scale to the same differences in $Q$ and $U$ ($\Delta Q=\Delta U$ in plot unit lengths).}
\label{f:quB}
\end{figure*}

\begin{figure*}\centering
\includegraphics[clip,trim=5mm 6mm 3mm 1mm,scale=.9,angle=90]{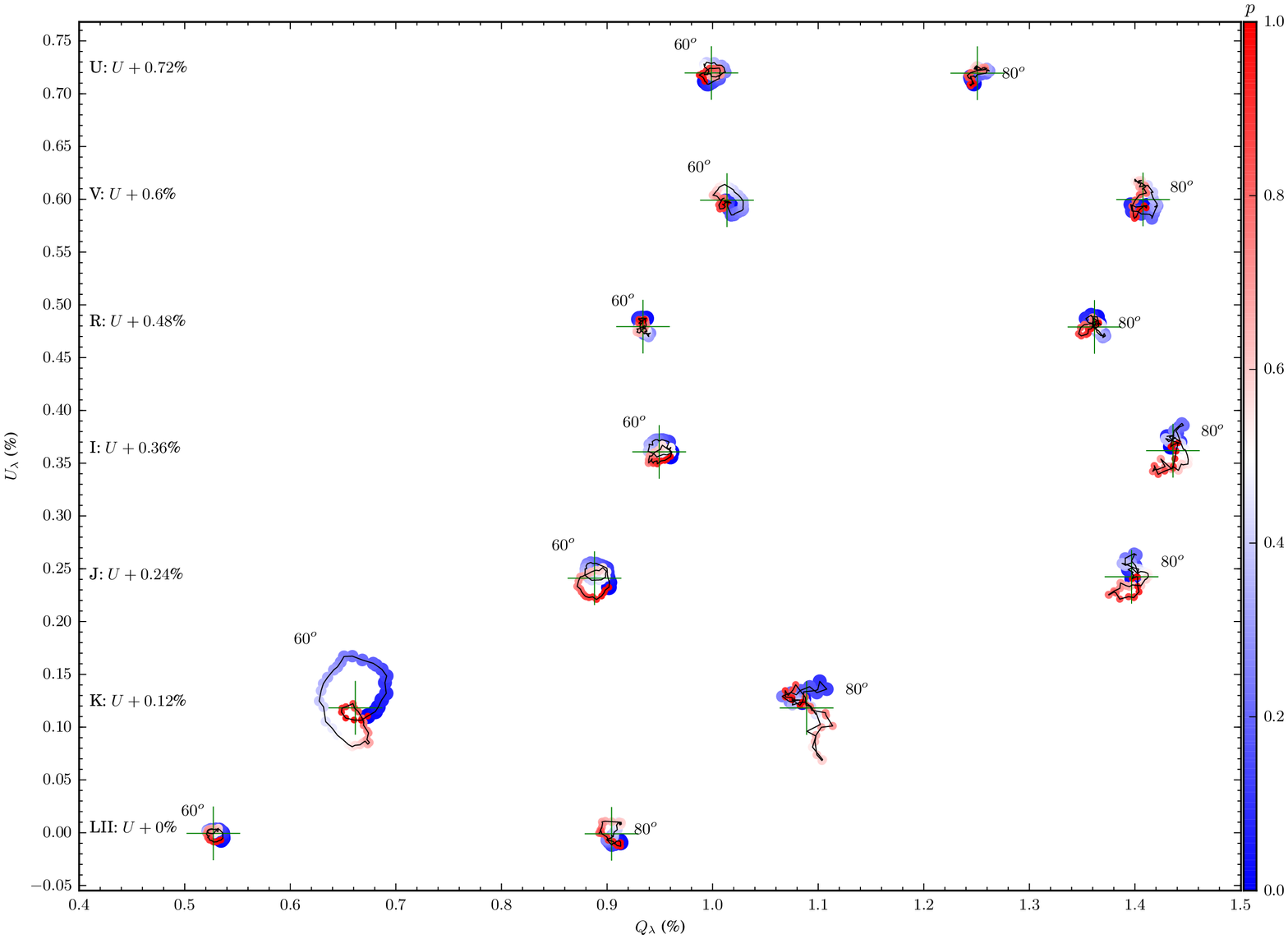}
\caption{Same as \fgr\ref{f:quB} at $i=60$ and $80\degr$.}\label{f:quM}
\end{figure*}

The description that follows is based on the discussion of \App\ref{s:how}. A series of $\QU$ diagrams was prepared for various spectral bands and inclinations (\fgrs\ref{f:quB} and \ref{f:quM}). The plotted diameter of each $\uqp$ point decreases with phase, so that it can be easy to trace the time evolution. This plotting feature also facilitates comparison of different bands at the same phase.

Let the point $\text{C}\equiv(\mean{Q},\mean{U})$ be defined as the central point of a $\QU$ cycle (C is depicted with a cross in each $\QU$ loop of \fgrs\ref{f:quB} and \ref{f:quM}). Then its coordinates in a $\QU$ diagram give the mean values of the $\uqp$ points for each band at the specified inclination.
That is similar (but not exactly; see \App\ref{s:how}) to the logic of ``distance from origin'' as a way to represent the intrinsic polarisation of a sample of stars in \fgro1 of \citet[ultraviolet+yellow polarimetry]{Serk70}. The same assumption is followed in the $\QU$ plots of \cite{McDa99}, where a dotted ellipse centred on the mean value represents the standard deviation.

At a first glance, two points can be made. First, the mean value of the $U$ polarisation component remains constant and equal to zero for any band regardless of inclination (see \fgr\ref{f:fqu}c and discussion about the PA in \Sm\ref{s:wd}). Second, the mean $Q$ increases with $i$ for any passband, which signifies that the mean polarisation $\mean{P}$ increases with $i$ (see \fgrs\ref{f:fqu}b, \ref{f:fqu}d and \ref{f:smean}).
All pole-on views show circular $\QU$ sequencies centred at the origin $\text{O}(0,0)$ of the coordinate system, which hints at maximal PTP differences of the PA across abrupt angle reversals (\fgrs\ref{f:varpa} and \ref{f:stpa}; \App\ref{s:how}). This is in agreement with the observations of \cite{McDa99}: The nearly pole-on star $\chi$ Oph has low polarisation and its time-dependent $\uqp$ points seem evenly scattered around a central point, showing no preferential mean PA.
The elongation of a time-dependent $\QU$ diagram towards higher inclinations hints at a preferential polarisation direction (PA; see \citealt{McDa99}). Note that near-circular and elliptical temporal $\QU$ diagrams have been observed by \cite{YuEv98} for a number of stars at different evolutionary stages.

From the decreasing radius of each plotted point in time, we see that, if the star is seen at $i=0\degr$, in fact there are two circular loops of equal sizes per orbital period through any passband. The loops are always counter-clockwise, i.e.\ they follow the orbital movement in the binary system (see the companion's movement in the pole-on snapshots of density in \fgro4 of \citetalias{PCO15}).
In particular, the \mbox{$(Q,U)$} points progress in the direction of movement of the disc gas (prograde motion), following the azimuthal structure of the disc. This might be an additional indication of the chirality of the disc observables, along with the sign of the mean $\log(\text{V/R})$, which is able to indicate the rotational direction of the star (see \So4.3.1 of \citetalias{PaFa16}).
From these facts it becomes obvious that in studies of temporal variations of polarisation it is important to show the time flow of the $\uqp$ points in a $\QU$ diagram, as in \fgro4 (third panel) of \cite{ReBl18}. If the time flow is not shown, then it is not possible to see if the $\uqp$ points change almost linearly with time, or if their apparent scattering in fact is a periodical variation around a mean value.

The two loops per period are seen at larger inclinations as well, but their sizes are not equal any more and they diverge from a circular shape. Two unequal and nearly circular loops indicate that a $P=f(p)$ plot can be approximated with two sinusoids of different amplitudes per orbital cycle.
So for the part of the cycle that the smaller-amplitude sinusoid dominates the variation of $P$, the arm that passes from the line of sight is too tenuous to produce a higher polarisation level. This characteristic is less pronounced at low inclinations, while it is more evident for infrared bands at higher inclinations (\fgrs\ref{f:quB} and \ref{f:quM}).

The PA is inferred at each point of a $\QU$ diagram from \eqt\eqref{e:PA}, and lies in the range $[-90,90]\degr$ (see \fgr\ref{f:geom}b and \App\ref{s:how}). Steep PA reversals can be seen in the nearly circular $\QU$ loops of \fgr\ref{f:quB}, better illustrated at \mbox{$i=10\degr$} for the U band (\fgr\ref{f:p}, with two abrupt angle reversals per cycle) or at $i=10,20\degr$ for the K band (\fgr\ref{f:p22}, with two and one abrupt angle reversals per cycle, respectively).
Withal, there are at least two angle reversals per orbital cycle for any passband and any inclination, but at low inclinations we may also see steep angle reversals.

The area enclosed in a $\QU$ loop of a single orbital period is related to the variance $\vari$ of polarisation (for a detailed analysis of this statement, see \App\ref{s:how}). The variability of $P$ can be qualified by how much the shape of the loop diverges from circular. Circularity is ascertained when the tangent on any infinitesimal part of the $\QU$ curve is vertical to the line that connects it to the central point C.
This pattern is evident on the nearly circular $\QU$ loops of all bands at $i=0\degr$, and diminishes at higher inclinations, especially for visible passbands. At larger inclinations, more than one intersections on the curve of a $\QU$ loop start to appear for all passbands. The intersections signify elimination of the area enclosed by the $\QU$ curve, so that it becomes smaller than $\upi\vari$.
The degeneracy of one of the circular loops to a smaller non-circular one conveys the occultation of the disk part that lies behind the star, especially its inner regions, where most of the scattered flux comes from. This symmetry-breaking effect is more pronounced at higher inclinations, where larger parts of the disc are concealed by the star.

The temporal variations are caused by the rotation of the disc and consequently by the periodic oscillations of overdense and underdense regions, as the spiral arms switch one after another on the line of sight.
The structure of the \mbox{$\QU$} loops is complex because of the changing part of the disc being on the observer's line of sight at each moment, accompanied by a wavelength-dependent opacity in a continuous interplay between absorption and scattering. The $\QU$ diagrams of \fgrs\ref{f:quB} and \ref{f:quM} confirm the statistics depicted in \fgrs\ref{f:smean}, \ref{f:sgen} and \ref{f:varpa}.
For instance, many intersections in visible bands veil the variability of polarisation already from $i=40\degr$. In addition, while polarisation in the infrared band K is smaller (i.e.\ smaller values of $Q$) than in other bands at intermediate inclinations (\fgr\ref{f:smean}, bottom), its variability is important even at $i=60\degr$ (relatively large variance -- larger area inside the $\QU$ curve), as indicated also from \fgr\ref{f:sgen} (bottom).

Parametric $\QU$ diagrams in which the changing parameter is the wavelength can be described in a similar way as temporal $\QU$ diagrams. In case of line polarisation the red continuum would be equal to the blue continuum in $Q$ and $U$, so there will be a closed $\QU$ loop across the line, whose shape can be explained as periodic temporal $\QU$ loops.

\begin{figure*}\centering
\includegraphics[clip,trim=2mm 11.75mm 1mm 1mm,scale=.9]{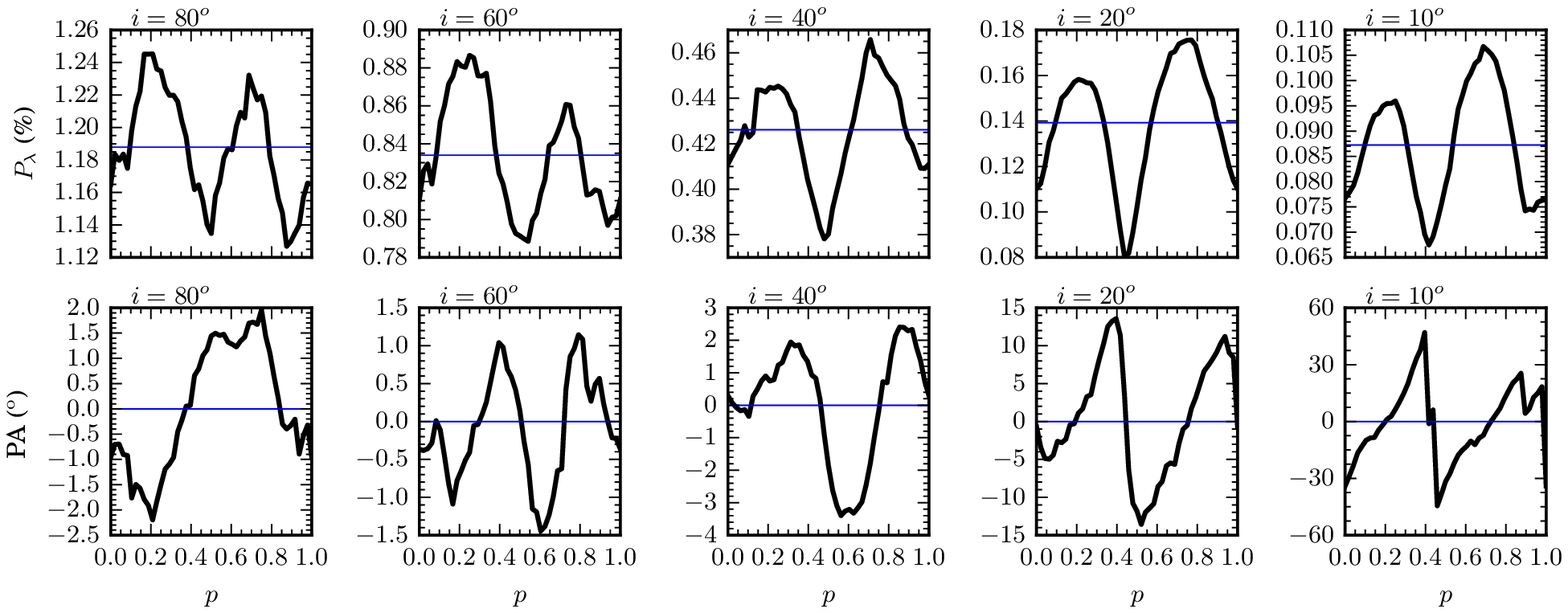}
\includegraphics[clip,trim=1.5mm 3mm 1mm 0mm,scale=.9]{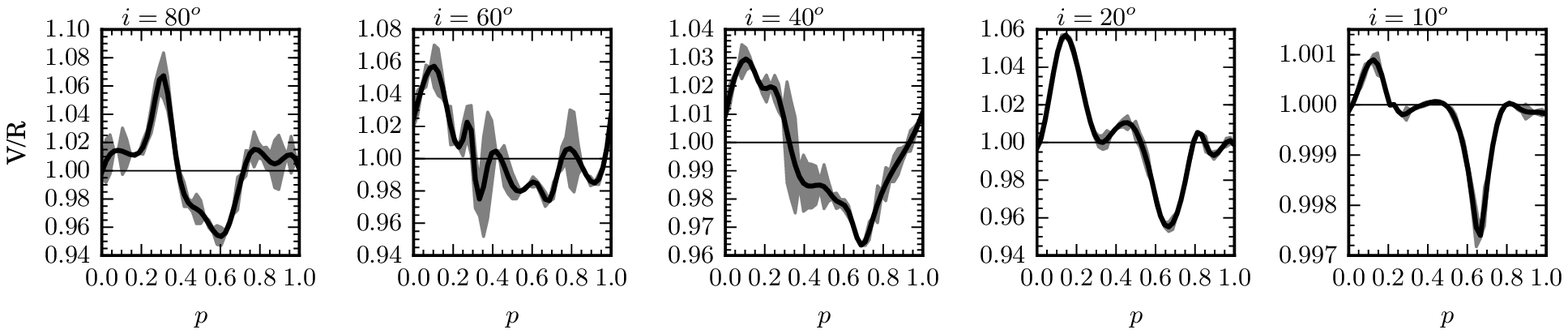}
\caption{The polarisation level (top) and PA (middle row) averaged over a narrow band around $\Ha$, and the $\Ha$ V/R ratio (bottom) as functions of the orbital phase at various inclinations, labelled on the top of each panel. The horizontal lines mark the mean polarisation (top), the zero-level in PA (middle row) and V/R=1 (bottom).
The V/R curves have been smoothed with splines, and the shaded area around each curve shows the residuals from the spline fitting. The fact that the spline residuals increase around V/R=1 is indicative of the velocity vectors being mixed around the line of sight during these phases, both in values and directions.}\label{f:VR}
\end{figure*}

\section{Discussion}\label{s:dis}
\subsection{Correlation of polarisation and line profiles}
\cite{McDa99} hypothesized a correlation between polarisation and V/R ratio in non-axisymmetric discs. \cite{McDa00} observed the B-band polarisation and the V/R ratio of $\Ha$ for the Be shell stars $\zeta$ Tau and 48~Lib. Considering an one-armed density wave in the disc, \citeauthor{McDa00} assumed that the two observables are locked to the rotational period of the spiral arm.
The V/R peak ratio was fitted with a period twice as long as that of polarisation cycles, but the selected period ratio 2:1 is of uncertain accuracy given the large scatter and scarcity of data (probably connected to the intervention of variabilities of other origins, as noted by \citeauthor{McDa00}).

The one or two peaks in polarisation and PA are reminiscent of the variability of the V/R ratio in \citetalias{PaFa16}. Our present calculations for the two-armed Be disc in a binary system confirm the connection between variations of V/R ratio and polarisation (\fgr\ref{f:VR}).
In terms of the V/R ratio, the variability amplitude does not increase by much with increasing inclination for the current set of system parameters, but there are still one or two blue-to-red transitions every period. At high inclinations, the highest V/R ratio seems to coincide with the highest polarisation levels, but there is always some phase difference. At low inclinations, one of the two polarisation maxima coincides with the lower value of the V/R ratio.

A connection between the V/R ratio and PA is more evident. In particular, around the phases where the PA changes sign the V/R ratio is about 1. This is true for all inclinations studied.
The V/R=1 condition is valid for a relatively extended range in phase, signifying that blue-to-red transitions occur for as long as velocities of mixed value and direction pass through the line of sight \citepalias{PaFa16}. Moreover, V/R=1 can be thought of as a symmetry indicator, meaning that blue velocities cancel out the red ones.

The two peaks per orbital cycle do not necessarily mean a variability frequency twice the orbital frequency. First, the two peaks may degrade to one (in inclinations far from pole-on), and this was seen in certain cases also in the study of the V/R ratio in \citetalias{PaFa16}. Second, while the variability frequency is equal to the orbital frequency, there is a phase difference $\Delta p$ between the two peaks (if more than one).
If $\Delta p=0.5$, then the variability frequency might be mistakenly thought of as double of the orbital frequency. A similar situation might occur in the V/R ratio: a variability frequency equal to the orbital frequency, and two (unequal) peaks per period that come with a phase lag that may or may not be equal to 0.5.

We note that while it is relatively easy to detect disc variability for edge-on discs through changes in the photometric flux and polarisation (\fgr\ref{f:sgen}) or spectroscopic features (\fgr\ref{f:VR}, bottom), these observables can hardly show variations for nearly pole-on discs. In the latter case, rapid changes in the PA might be the only means to detect a variable disc (\fgr\ref{f:varpa}), provided an adequate temporal resolution.

\subsection{Distinguishing classical Be stars based on their polarisation signature}
In a (UBVRI) polarimetric study of clusters with large populations of B-type stars, \cite{WiBj07} classified the sources depending on their likelihood of being classical Be stars. The classification scheme of \citeauthor{WiBj07} is quite elaborate, but the effect of inclination angle occasionally was underestimated.

The characteristic that was considered the most crucial indication of a classical Be star is the sawtooth wavelength-dependent shape of the polarisation profile. However, our \fgr\ref{f:fqu}(d) shows that the Balmer and Paschen jumps might be veiled at very small and very large inclinations. Thus \citeauthor{WiBj07} correctly did not rule out objects without a sawtooth-type polarisation.
The authors explained this exception only in terms of density, i.e.\ a disc of very low density would anyway have a small scattering opacity, thus it would be unable to rescatter light and raise the polarisation level longwards the limits of hydrogen series. In addition, they added some likelihood of the object being a pole-on Be star, in which case it would be minimally polarised (\fgr\ref{f:smean}).
As shown here and in other works (see \Sm\ref{s:wd} for a discussion), a Be star with high-density disc seen at {90\degr} might also not have polarisation with a pronounced sawtooth-type shape.

Another feature that was considered by \cite{WiBj07} as able to make the object under study more likely to be a classical Be star is a slight dependence of PA on wavelength. None the less, our \fgr\ref{f:fqu}(e) shows that PA does not really change with wavelength, other than some discontinuities from recombination lines.
In fact the PA (or the mean PA, for periodic variations) is not expected to change for a given star, because the orientation of the disc remains constant \citep{QuBj97}. Anyhow, \citeauthor{WiBj07} note that a wavelength dependence of the PA might also indicate B[e] supergiants surrounded by thin dusty discs, as shown in optical polarimetry by \citet{Maga92}.

Angle reversals is another feature explored by \cite{WiBj07}. The authors regard angle reversals as a probable indication of the non-Be nature of the source, and instead consider it a characteristic of dusty environments. However, as depicted in \fgrs\ref{f:p}-\ref{f:pl} and \ref{f:VR}, angle reversals do emerge with time from pure Thomson scattering in dense non-symmetric discs, while \fgr\ref{f:fqu}(e) shows that angle reversals occur also as a function of wavelength.
The existence of PA reversals does not appear sufficient to constrain the nature of the observed system or to decide whether its atmosphere is dusty or not, but is rather indicative of asymmetries.

In addition to broad-band polarimetry, narrow-band spectropolarimetric observations in $\Ha$ and $\Hb$ by \cite{MlCl79} revealed that neither supergiants nor Wolf-Rayet systems show significant depolarisation levels, but only Be stars.
This might be due to the fact that such systems are more spherically symmetric and/or more dusty. Dust may induce a circular polarisation component that is typically considered zero in hot gaseous environments. However, as already mentioned (\Sms\ref{s:elp} and \ref{s:pd}), GG Car is a dusty B[e] supergiant that does show line depolarisation.

We further note that higher polarisation levels (at visible wavelengths) were found by \citeauthor{MlCl79} in Be stars later reported as having orientations closer to edge-on. One of those Be stars, $\eta$ Cen, showed increased polarisation at the time of observations. To our knowledge no recently published polarimetric survey has included $\eta$~Cen as a target.
It would thus be worthwhile to investigate this star polarimetrically again, especially because its recent photometric study showed permanent but variable mass loss driven by non-radial pulsations \citep{Brite1}. Spectropolarimetry might be used to infer the geometry of the process of mass injection from the star to the disc.

\subsection{Model applications and limitations}\label{s:lim}
Throughout this text, we have compared our results to observations not only of Be stars in coplanar circular binaries with discs of two-armed spiral structures (which is what was actually modelled), but of various stars that may show variations in observational features indicative of discs with azimuthal asymmetries (e.g.\ $\beta$ Lyr, Be stars with one-armed discs, B[e] supergiants).
The similarities of such observations suggest that disc variations may be ambiguous in defining the physical mechanism that causes them, but they do reveal the geometrical asymmetries that result from this mechanism.
Hence, the properties of non-symmetric discs, as those shaped due to the tidal forces from a binary companion, can help explain the periodical (and not only) behaviour of different classes of stars. Similarly, in a recent study on double contact binaries, \cite{Wils18} notes that his results for the eclipses in accretion-decretion systems can also explain the behaviour of other types of stars.

The target of this study is the periodical variations in observables that emerge from azimuthally non-symmetric steady rotating discs. Given a period $\Porb$ of the variation, periodical variations imply that the value of a quantity $M$ satisfies $M(t+\Porb)=M(t)\ \forall t$.
The parametric plots of any two quantities (such as the two Stokes parametres $Q$ and $U$) would be closed loops. However, this is rarely what is actually observed, since entangled variabilities can disturb a perfect periodicity. As a result, no perfectly closed $\QU$ loops really exist, and temporal monotonic trends would prevent the value of $M$ from repeating every $\Porb$.

Of course there might be cases where other mechanisms affect the disc too little, and a periodicity can be seen. For instance, in some binaries a combination of disc truncation and high mass ejection rates can effectively lead to a nearly stable disc, at least in the region inside truncation, where the density is higher. Potential examples of such cases may be the Be stars $\phi$ Per and 59~Cyg.
Both are nearly circular binaries with sdO companions and orbital periods of 127 and 28 d, respectively, seen at a large inclination \citep{PePe13}.
There are HPOL polarimetric data\footnote{\url{http://www.sal.wisc.edu/~meade/beatlas/objects.html}} in the V band for these stars, but the temporal resolution is too coarse to affirm the existence of closed $\QU$ loops, especially since they are expected to be noisy (many intersections) due to the high inclination.

In \Sm\ref{s:qU} it was emphasized that temporal $\QU$ diagrams should be accompanied by a description of the time evolution of the $\uqp$ points. Simple linear least-square fits of the coordinates of the $\uqp$ points miss the temporal information.
The importance of tracing the time evolution becomes obvious in works as that of \cite{HaGu84} on the Be star $\omega$ Ori, who describe a monotonic change from $(Q_1,U_1)$ to $(Q_2,U_2)$ and then back to $(Q_1,U_1)$, this time in a backward direction. This behaviour can still be approximately characterised as a closed $\QU$ loop, similar to that seen e.g.\ for the K band at $i=80\degr$ in \fgr\ref{f:quM}.
The scatter around the linear least-square fit of similar $\QU$ plots might be alternatively seen as intersections (see \App\ref{s:how}). Taking into account the variations of simultaneous photometry, \citeauthor{HaGu84} qualify this event as a mass loss episode with the turning point marking the episode's ceasing and the beginning of the system's relaxation.

As in \cite{Wils18}, perfectly optically thin discs are not targeted. A perfectly optically thin disc would probably show spectroscopic variations mainly in EW. On the contrary, optically thick or semi-transparent discs can show double-peaked profiles at orientations close to edge-on, with peak heights higher at larger inclinations. At a range of intermediate inclinations the line profiles would be flat-topped \citepalias{PaFa16}.
After this inclination range the line profiles become single-peaked, with the peak height increasing as we approach a pole-on orientation. The optical thickness is basically determined by the disc density and the amount of disc volume that the gas has to cross in order to reach the observer.

\section{Conclusions}\label{s:con}
The primary aim of this work was to theoretically verify a connection between V/R line ratios and polarimetric observables for coplanar circular Be binaries. This target was met successfully, with its most direct manifestation being that the condition \mbox{$\text{V/R}\simeq1$} generally coincides with phases where the PA changes sign.

Polarigenic observables of binary Be discs generally show a high peak twice per cycle, each associated with either one of the spiral arms passing from the line of sight. For both $P$ and PA there are two maxima per period, generally unequal and with a phase difference between them that can be $\Delta p\ne 0.5$. The variability in both observables increases at orientations closer to pole-on, it is more evident in PA, and is maintained even at large inclinations for infrared bands.
In particular, the polarisation level of infrared passbands, although a little smaller in value, shows more pronounced variability at any inclination. It would therefore be interesting to confirm our model's results by polarimetric observations outside the visible spectrum, as proposed by \cite{BjorK0} and \cite{WiBj07}.

The PA in general can show periodic variability more significant than continuum flux or polarisation, especially if the star is seen at low inclinations. This is particularly important for pole-on discs, where the variability in other observables diminishes. The PA might change very rapidly, the extreme being steep angle reversals. Variability in PA is particularly important for non-axisymmetric discs, as those formed by binary interaction or by global disc oscillations.
As noted by \cite{HaJo13b} other polarimetric features may have qualitatively similar behaviour in discs dynamically changing in different ways. For instance, in terms of the BJV diagram (Balmer jump against V-band polarisation, \citealt{WiBj07}), the periodic variability in growing/dissipating discs is similar to non-axisymmetric discs with spiral arms.
On the contrary, the PA is nearly constant and parallel to the symmetry axis in axisymmetric discs, no matter whether they are growing or shrinking by mass load or removal (\citealt{QuBj97}). Therefore variability in PA implicitely indicates non-axisymmetric structures.

We also explored narrow bands around recombination lines typical in Be discs. The drop in the polarisation level at wavelengths of hydrogen lines must be a complex function of the disc properties and the inclination angle. There is an apparent (anti)correlation between an emission line profile and depolarisation, with the polarisation profile relative to continuum of a line having a shape similar to the corresponding relative flux, but reversed.
An investigation of $\QU$ plots across the wavelength regions of hydrogen lines will probably give more insight to further information.

An important outcome of this paper is the decipherment of time-dependent $\QU$ diagrams. It was found that the area encircled by a $\QU$ curve over a variability cycle (regardless of origin) is connected to the variance of polarisation. The shape of a $\QU$ loop can be used for the qualification of the periodicity of the variability, and the divergence of the shape from circular is a measure for the non-sinusoidality of the variation of $P^2$.
We also demonstrated why temporal $\QU$ diagrams should include information on the time evolution, and that linear least-square fits of the $\uqp$ points are not always the ideal way to describe a $\QU$ plot.

An exploration of the system parameters is on-going. Preliminary results for coplanar circular binaries of other parameter sets indicate that the slight photometric variability over the orbital cycle will increase in amplitude for closer binaries, in accord with line-profile variability. Closer binaries also show higher variability in PA than wider binaries, while linear polarization is lower but it varies in a similar manner.
Lower density discs, as those that would form in higher viscosity regimes \citepalias{PCO15}, would be less polarised. The variability amplitudes for both linear polarisation and PA are also lower in more tenuous discs.

Modelling studies of binary Be discs as the present one can help understand additional phenomena, since their observational consequences can be similar. As linear polarisation arises in the region with the greatest number of scatterers, it probes the inner parts of the disc, which are denser and even in direct interchange of mass and angular momentum with the photosphere.
It follows that parallel spectroscopic and polarimetric observations at high temporal and spectral resolution might make it possible to remove uncertainties on the state of the gas, as well as on disc formation and evolution.

\paragraph*{Acknowledgements.} DP thanks the referee, John Wisniewski, for the encouraging and thoughtful remarks that improved the quality of this paper. This work made use of the computing facilities of the Group of Applied Geophysics (Observat\'orio Nacional, Brazil) and of the Laboratory of Astroinformatics (IAG/USP; financed by the brazilian agency FAPESP through grant \mbox{2009/54006-4}).
DP acknowledges financial support from Conselho Nacional de Desenvolvimento Cient\'ifico e Tecnol\'ogico (Brazil) through grant 300235/2017-8. DMF acknowledges FAPESP \mbox{(2016/16844-1)}. This research has made use of the SVO Filter Profile Service\footnote{\url{http://svo2.cab.inta-csic.es/theory/fps/}} supported by the Spanish MINECO through grant AyA2014-55216.

\bibliography{biblio}

\begin{thebibliography}{}
\makeatletter
\relax
\def\mn@urlcharsother{\let\do\@makeother \do\$\do\&\do\#\do\^\do\_\do\%\do\~}
\def\mn@doi{\begingroup\mn@urlcharsother \@ifnextchar [ {\mn@doi@}
  {\mn@doi@[]}}
\def\mn@doi@[#1]#2{\def\@tempa{#1}\ifx\@tempa\@empty \href
  {http://dx.doi.org/#2} {doi:#2}\else \href {http://dx.doi.org/#2} {#1}\fi
  \endgroup}
\def\mn@eprint#1#2{\mn@eprint@#1:#2::\@nil}
\def\mn@eprint@arXiv#1{\href {http://arxiv.org/abs/#1} {{\tt arXiv:#1}}}
\def\mn@eprint@dblp#1{\href {http://dblp.uni-trier.de/rec/bibtex/#1.xml}
  {dblp:#1}}
\def\mn@eprint@#1:#2:#3:#4\@nil{\def\@tempa {#1}\def\@tempb {#2}\def\@tempc
  {#3}\ifx \@tempc \@empty \let \@tempc \@tempb \let \@tempb \@tempa \fi \ifx
  \@tempb \@empty \def\@tempb {arXiv}\fi \@ifundefined
  {mn@eprint@\@tempb}{\@tempb:\@tempc}{\expandafter \expandafter \csname
  mn@eprint@\@tempb\endcsname \expandafter{\@tempc}}}

\bibitem[\protect\citeauthoryear{Baade et~al.,}{Baade et~al.}{2016}]{Brite1}
Baade D.,  et~al., 2016, \mn@doi [\aap] {10.1051/0004-6361/201528026}, 588, A56

\bibitem[\protect\citeauthoryear{Bidelman}{Bidelman}{1976}]{Bide76}
Bidelman W.~P.,  1976, in Slettebak A.,  ed.,  IAU Symp Vol. 70, Be and shell
  stars. D.~Reidel, Dordrecht, p.~457

\bibitem[\protect\citeauthoryear{Bjorkman}{Bjorkman}{2000}]{BjorK0}
Bjorkman K.~S.,  2000, in \cite{SmHe00}, p.~384

\bibitem[\protect\citeauthoryear{Bjorkman et~al.,}{Bjorkman
  et~al.}{1991}]{BjNo91}
Bjorkman K.~S.,  et~al., 1991, \apj, 383, L67

\bibitem[\protect\citeauthoryear{Bjorkman, Miroshnichenko, McDavid  \&
  Pogrosheva}{Bjorkman et~al.}{2002}]{BjMi02}
Bjorkman K.~S.,  Miroshnichenko A.~S.,  McDavid D.,   Pogrosheva T.~M.,  2002,
  \mn@doi [\apj] {10.1086/340751}, 573, 812

\bibitem[\protect\citeauthoryear{Bohren \& Huffman}{Bohren \&
  Huffman}{1983}]{BoHu83}
Bohren C.~F.,  Huffman D.~R.,  1983, {Absorption and scattering of light by
  small particles}.
Interscience, Wiley, NY (\S3.2)

\bibitem[\protect\citeauthoryear{Borges~Fernandes}{Borges~Fernandes}{2010}]{Bofe10}
Borges~Fernandes M.,  2010, in Rivinius T.,  Cur\'e M.,  eds,
  Rev.~Mex.~Astron.~Astrofis. Conf.~Ser. Vol. 38, {The interferometric view of
  hot stars}. p.~98

\bibitem[\protect\citeauthoryear{Brown \& McLean}{Brown \&
  McLean}{1977}]{BrMl77}
Brown J.~C.,  McLean I.~S.,  1977, \aap, 57, 141

\bibitem[\protect\citeauthoryear{Carciofi \& Bjorkman}{Carciofi \&
  Bjorkman}{2006}]{CaBj06}
Carciofi A.~C.,  Bjorkman J.~E.,  2006, \mn@doi [\apj] {10.1086/499483}, 639,
  1081

\bibitem[\protect\citeauthoryear{Chandrasekhar}{Chandrasekhar}{1946}]{Chan46a}
Chandrasekhar S.,  1946, \apj, 103, 351

\bibitem[\protect\citeauthoryear{Chandrasekhar}{Chandrasekhar}{1960}]{Chan60}
Chandrasekhar S.,  1960, {Radiative transfer}.
Dover, New York (\S15.1)

\bibitem[\protect\citeauthoryear{Clark, Tarasov  \& Panko}{Clark
  et~al.}{2003}]{ClTa03}
Clark J.~S.,  Tarasov A.~E.,   Panko E.~A.,  2003, \mn@doi [\aap]
  {10.1051/0004-6361:20030248}, 403, 239

\bibitem[\protect\citeauthoryear{Coyne}{Coyne}{1970}]{Coyn70}
Coyne G.~V.,  1970, Ricerce Astronomiche, 8, 85

\bibitem[\protect\citeauthoryear{Coyne}{Coyne}{1971}]{Coyn71}
Coyne G.~V.,  1971, Ricerce Astronomiche, 8, 201

\bibitem[\protect\citeauthoryear{Coyne \& Kruszewski}{Coyne \&
  Kruszewski}{1969}]{CoKr69}
Coyne G.~V.,  Kruszewski A.,  1969, \aj, 74, 528

\bibitem[\protect\citeauthoryear{Delaa et~al.,}{Delaa et~al.}{2011}]{DeSt11}
Delaa O.,  et~al., 2011, \mn@doi [\aap] {10.1051/0004-6361/201015639}, 529, A87

\bibitem[\protect\citeauthoryear{Halonen \& Jones}{Halonen \&
  Jones}{2013a}]{HaJo13b}
Halonen R.~J.,  Jones C.~E.,  2013a, \apjs, 208, 3

\bibitem[\protect\citeauthoryear{Halonen \& Jones}{Halonen \&
  Jones}{2013b}]{HaJo13a}
Halonen R.~J.,  Jones C.~E.,  2013b, \apj, 765, 17

\bibitem[\protect\citeauthoryear{Harmanec}{Harmanec}{1988}]{Harm88}
Harmanec P.,  1988, Bull.~Astr.~Inst.\ Czechoslovakia, 39, 329

\bibitem[\protect\citeauthoryear{Harrington \& Kuhn}{Harrington \&
  Kuhn}{2009}]{HaKu09}
Harrington D.~M.,  Kuhn J.~R.,  2009, \mn@doi [\apjs]
  {10.1088/0067-0049/180/1/138}, 180, 138

\bibitem[\protect\citeauthoryear{Hayes \& Guinan}{Hayes \&
  Guinan}{1984}]{HaGu84}
Hayes D.~P.,  Guinan E.~F.,  1984, \mn@doi [\apj] {10.1086/161938}, 279, 721

\bibitem[\protect\citeauthoryear{Hern\'andez, L\'opez, Sahade  \&
  Thackeray}{Hern\'andez et~al.}{1981}]{HeLo81}
Hern\'andez C.~A.,  L\'opez L.,  Sahade J.,   Thackeray A.~D.,  1981, \mn@doi
  [\pasp] {10.1086/130920}, 93, 747

\bibitem[\protect\citeauthoryear{Lamers, Zickgraf, de Winter, Houziaux  \&
  Zorec}{Lamers et~al.}{1998}]{LaZi98}
Lamers H.~J.~G.~L.~M.,  Zickgraf F.-J.,  de Winter D.,  Houziaux L.,   Zorec
  J.,  1998, \aap, \href {http://adsabs.harvard.edu/abs/1998A%26A...340..117L}
  {340, 117}

\bibitem[\protect\citeauthoryear{Lee, Saio  \& Osaki}{Lee
  et~al.}{1991}]{LeOs91}
Lee U.,  Saio H.,   Osaki Y.,  1991, \mnras, 250, 432

\bibitem[\protect\citeauthoryear{Magalh\~aes}{Magalh\~aes}{1992}]{Maga92}
Magalh\~aes A.~M.,  1992, \mn@doi [\apj] {10.1086/171856}, 398, 286

\bibitem[\protect\citeauthoryear{Marchiano, Brandi, Muratore, Quiroga, Ferrer
  \& Garc\'ia}{Marchiano et~al.}{2012}]{MaBr12}
Marchiano P.,  Brandi E.,  Muratore M.~F.,  Quiroga C.,  Ferrer O.~E.,
  Garc\'ia L.~G.,  2012, \aap, 540, A91

\bibitem[\protect\citeauthoryear{McDavid}{McDavid}{1999}]{McDa99}
McDavid D.,  1999, \mn@doi [\pasp] {10.1086/316349}, 111, 494

\bibitem[\protect\citeauthoryear{McDavid, Bjorkman, Bjorkman  \&
  Okazaki}{McDavid et~al.}{2000}]{McDa00}
McDavid D.,  Bjorkman K.~S.,  Bjorkman J.~E.,   Okazaki A.~T.,  2000, in
  \cite{SmHe00}, p.~460

\bibitem[\protect\citeauthoryear{McLean \& Clarke}{McLean \&
  Clarke}{1979}]{MlCl79}
McLean I.~S.,  Clarke D.,  1979, \mn@doi [\mnras] {10.1093/mnras/186.2.245},
  186, 245

\bibitem[\protect\citeauthoryear{Mihalas}{Mihalas}{1967}]{Miha67}
Mihalas D.,  1967, \mn@doi [\apj] {10.1086/149239}, 149, 169

\bibitem[\protect\citeauthoryear{Nordsieck, Babler, Bjorkman, Meade,
  Schulte-Ladbeck  \& Taylor}{Nordsieck et~al.}{1992}]{NoBa92}
Nordsieck K.~H.,  Babler B.,  Bjorkman K.~S.,  Meade M.~R.,  Schulte-Ladbeck
  R.~E.,   Taylor M.~J.,  1992, in Drissen L.,  Leitherer C.,   Nota A.,  eds,
  ASP Conf.\ Ser. Vol. 22, {Nonisotropic and variable outflows from stars}.
  Astron.\ Soc.\ Pac., San Fransisco, p.~114

\bibitem[\protect\citeauthoryear{Okazaki}{Okazaki}{1991}]{Okaz91}
Okazaki A.~T.,  1991, \pasj, 43, 75

\bibitem[\protect\citeauthoryear{Okazaki}{Okazaki}{1997}]{Okaz97a}
Okazaki A.~T.,  1997, \aap, 318, 548

\bibitem[\protect\citeauthoryear{Okazaki, Bate, Ogilvie  \& Pringle}{Okazaki
  et~al.}{2002}]{OkBa02}
Okazaki A.~T.,  Bate M.~R.,  Ogilvie G.~I.,   Pringle J.~E.,  2002, \mn@doi
  [\mnras] {10.1046/j.1365-8711.2002.05960.x}, 337, 967

\bibitem[\protect\citeauthoryear{Panoglou, Carciofi, Vieira, Cyr, Jones,
  Okazaki  \& Rivinius}{Panoglou et~al.}{2016}]{PCO15}
Panoglou D.,  Carciofi A.~C.,  Vieira R.~G.,  Cyr I.~H.,  Jones C.~E.,  Okazaki
  A.~T.,   Rivinius T.,  2016, \mn@doi [\mnras] {10.1093/mnras/stw1508}, 461,
  2616. Paper~I

\bibitem[\protect\citeauthoryear{Panoglou, Faes, Carciofi, Okazaki, Baade,
  Rivinius  \& {Borges Fernandes}}{Panoglou et~al.}{2018}]{PaFa16}
Panoglou D.,  Faes D.~M.,  Carciofi A.~C.,  Okazaki A.~T.,  Baade D.,  Rivinius
  T.,   {Borges Fernandes} M.,  2018, \mn@doi [\mnras] {10.1093/mnras/stx2497},
  473, 3039. Paper II

\bibitem[\protect\citeauthoryear{Papaloizou \& Lin}{Papaloizou \&
  Lin}{1995}]{PaLi95}
Papaloizou J.~C.~B.,  Lin D.~N.~C.,  1995, \mn@doi [\araa]
  {10.1146/annurev.aa.33.090195.002445}, 33, 505

\bibitem[\protect\citeauthoryear{Papaloizou, Savonije  \& Henrichs}{Papaloizou
  et~al.}{1992}]{PaSa92}
Papaloizou J.~C.,  Savonije G.~J.,   Henrichs H.~F.,  1992, \aap, 265, L45

\bibitem[\protect\citeauthoryear{{Papoulis}}{{Papoulis}}{1965}]{Papo65}
{Papoulis} A.,  1965, {Probability, random variables and stochastic processes}.
McGraw-Hill, New York (\S6-9)

\bibitem[\protect\citeauthoryear{Pereyra, de Ara\'ujo, Magalh\~aes,
  Borges~Fernandes  \& Domiciano~de Souza}{Pereyra et~al.}{2009}]{PeAr09}
Pereyra A.,  de Ara\'ujo F.~X.,  Magalh\~aes A.~M.,  Borges~Fernandes M.,
  Domiciano~de Souza A.,  2009, \mn@doi [\aap] {10.1051/0004-6361/200913250},
  508, 1337

\bibitem[\protect\citeauthoryear{Peters, Pewett, Gies, Touhami  \&
  Grundstrom}{Peters et~al.}{2013}]{PePe13}
Peters G.~J.,  Pewett T.~D.,  Gies D.~R.,  Touhami Y.~N.,   Grundstrom E.~D.,
  2013, \mn@doi [\apj] {10.1088/0004-637X/765/1/2}, 765, 2

\bibitem[\protect\citeauthoryear{Poeckert \& Marlborough}{Poeckert \&
  Marlborough}{1976}]{PoMa76}
Poeckert R.,  Marlborough J.~M.,  1976, \mn@doi [\apj] {10.1086/154369}, 206,
  182

\bibitem[\protect\citeauthoryear{Poeckert \& Marlborough}{Poeckert \&
  Marlborough}{1978}]{PoMa78a}
Poeckert R.,  Marlborough J.~M.,  1978, \mn@doi [\apj] {10.1086/155984}, 220,
  940

\bibitem[\protect\citeauthoryear{Poeckert, Bastien  \& Landstreet}{Poeckert
  et~al.}{1979}]{PoBa79}
Poeckert R.,  Bastien P.,   Landstreet J.~D.,  1979, \mn@doi [\aj]
  {10.1086/112484}, 84, 812

\bibitem[\protect\citeauthoryear{Quirrenbach et~al.,}{Quirrenbach
  et~al.}{1997}]{QuBj97}
Quirrenbach A.,  et~al., 1997, \mn@doi [\apj] {10.1086/303854}, 479, 477

\bibitem[\protect\citeauthoryear{Reig}{Reig}{2011}]{Reig11}
Reig P.,  2011, \mn@doi [\apss] {10.1007/s10509-010-0575-8}, 332, 1

\bibitem[\protect\citeauthoryear{Reig \& Blinov}{Reig \& Blinov}{2018}]{ReBl18}
Reig P.,  Blinov D.,  2018, \mn@doi [\aap] {10.1051/0004-6361/201833649}, 619,
  A19

\bibitem[\protect\citeauthoryear{Riemann}{Riemann}{1867}]{Riem67}
Riemann B.,  1867, {Grundlagen f\"ur eine allgemeine Theorie der Functionen
  einer ver\"anderlichen complexen Gr\"osse}.
Verlag von Adalbert Rente, G\"ottingen

\bibitem[\protect\citeauthoryear{Serkowski}{Serkowski}{1970}]{Serk70}
Serkowski K.,  1970, \mn@doi [\apj] {10.1086/150497}, 160, 1107

\bibitem[\protect\citeauthoryear{Shakura \& Sunyaev}{Shakura \&
  Sunyaev}{1973}]{ShSu73}
Shakura N.~I.,  Sunyaev R.~A.,  1973, \aap, \href
  {http://adsabs.harvard.edu/abs/1973A%26A....24..337S} {24, 337}

\bibitem[\protect\citeauthoryear{Smith, Henrichs  \& Fabregat}{Smith
  et~al.}{2000}]{SmHe00}
Smith M.~A.,  Henrichs H.~F.,   Fabregat J.,  eds, 2000, The Be phenomenon in
  early-type stars  ASP Conf.\ Ser. Vol. 214.
Astron.\ Soc.\ Pac., San Fransisco

\bibitem[\protect\citeauthoryear{{\v Stefl}, Okazaki, Rivinius  \& Baade}{{\v
  Stefl} et~al.}{2007}]{StOk07}
{\v Stefl} S.,  Okazaki A.~T.,  Rivinius T.,   Baade D.,  2007, in {\v Stefl}
  S.,  Owocki S.~P.,   Okazaki A.~T.,  eds,  ASP Conf.\ Ser. Vol. 361, Active
  OB stars: Laboratories for stellar and circumstellar physics. Astron.\ Soc.\
  Pac., San Fransisco, p.~274

\bibitem[\protect\citeauthoryear{{\v Stefl} et~al.,}{{\v Stefl}
  et~al.}{2009}]{SRCl09}
{\v Stefl} S.,  et~al., 2009, \mn@doi [\aap] {10.1051/0004-6361/200811573},
  504, 929

\bibitem[\protect\citeauthoryear{Stokes}{Stokes}{1852}]{Sto852a}
Stokes G.~G.,  1852, Trans.\ Cambridge Phil.\ Soc., 9, 399

\bibitem[\protect\citeauthoryear{Vink, Drew, Harries  \& Oudmaijer}{Vink
  et~al.}{2002}]{ViDr02}
Vink J.~S.,  Drew J.~E.,  Harries T.~J.,   Oudmaijer R.~D.,  2002, \mn@doi
  [\mnras] {10.1046/j.1365-8711.2002.05920.x}, 337, 356

\bibitem[\protect\citeauthoryear{Vink, Harries  \& Drew}{Vink
  et~al.}{2005}]{ViHa05}
Vink J.~S.,  Harries T.~J.,   Drew J.~E.,  2005, \mn@doi [\aap]
  {10.1051/0004-6361:20041463}, 430, 213

\bibitem[\protect\citeauthoryear{Wilson}{Wilson}{2018}]{Wils18}
Wilson R.~E.,  2018, \mn@doi [\apj] {10.3847/1538-4357/aae6cc}, 869, 19

\bibitem[\protect\citeauthoryear{Wisniewski, Bjorkman  \&
  Magalh\~aes}{Wisniewski et~al.}{2003}]{WiBj03}
Wisniewski J.~P.,  Bjorkman K.~S.,   Magalh\~aes A.~M.,  2003, \mn@doi [\apjl]
  {10.1086/380500}, \href {http://adsabs.harvard.edu/abs/2003ApJ...598L..43W}
  {598, L43}

\bibitem[\protect\citeauthoryear{Wisniewski, Kowalski, Bjorkman, Bjorkman  \&
  Carciofi}{Wisniewski et~al.}{2007a}]{WiKo07}
Wisniewski J.~P.,  Kowalski A.~F.,  Bjorkman K.~S.,  Bjorkman J.~E.,   Carciofi
  A.~C.,  2007a, \mn@doi [\apjl] {10.1086/512123}, 656, L21

\bibitem[\protect\citeauthoryear{Wisniewski, Bjorkman, Magalh\~aes, Bjorkman,
  Meade  \& Pereyra}{Wisniewski et~al.}{2007b}]{WiBj07}
Wisniewski J.~P.,  Bjorkman K.~S.,  Magalh\~aes A.~M.,  Bjorkman J.~E.,  Meade
  M.~R.,   Pereyra A.,  2007b, \mn@doi [\apj] {10.1086/522293}, 671, 2040

\bibitem[\protect\citeauthoryear{Wisniewski, Draper, Bjorkman, Maede, Bjorkman
  \& Kowalski}{Wisniewski et~al.}{2010}]{WiDr10}
Wisniewski J.~P.,  Draper Z.~H.,  Bjorkman K.~S.,  Maede M.~R.,  Bjorkman
  J.~E.,   Kowalski A.~F.,  2010, \apj, 709, 1306

\bibitem[\protect\citeauthoryear{Wood, Bjorkman, Whitney  \& Code}{Wood
  et~al.}{1996a}]{WoBj96a}
Wood K.,  Bjorkman J.~E.,  Whitney B.~A.,   Code A.~D.,  1996a, \mn@doi [\apj]
  {10.1086/177105}, 461, 828

\bibitem[\protect\citeauthoryear{Wood, Bjorkman, Whitney  \& Code}{Wood
  et~al.}{1996b}]{WoBj96b}
Wood K.,  Bjorkman J.~E.,  Whitney B.~A.,   Code A.~D.,  1996b, \mn@doi [\apj]
  {10.1086/177106}, 461, 847

\bibitem[\protect\citeauthoryear{Yudin}{Yudin}{2001}]{Yudi01}
Yudin R.~V.,  2001, \mn@doi [\aap] {10.1051/0004-6361:20000577}, 368, 912

\bibitem[\protect\citeauthoryear{Yudin \& Evans}{Yudin \& Evans}{1998}]{YuEv98}
Yudin R.~V.,  Evans A.,  1998, \mn@doi [\aaps] {10.1051/aas:1998279}, \href
  {http://adsabs.harvard.edu/abs/1998A%26AS..131..401Y} {131, 401}

\bibitem[\protect\citeauthoryear{Zellner \& Serkowski}{Zellner \&
  Serkowski}{1972}]{ZeSe72}
Zellner B.~H.,  Serkowski K.,  1972, \mn@doi [\pasp] {10.1086/129343}, 84, 619

\makeatother
\end{thebibliography}
\appendix

\section{Geometrical and statistical properties of $\QU$ diagrams}
\label{s:how}

\begin{figure*}\centering
\includegraphics[clip,trim=1mm 3mm 2mm 2mm,scale=.9]{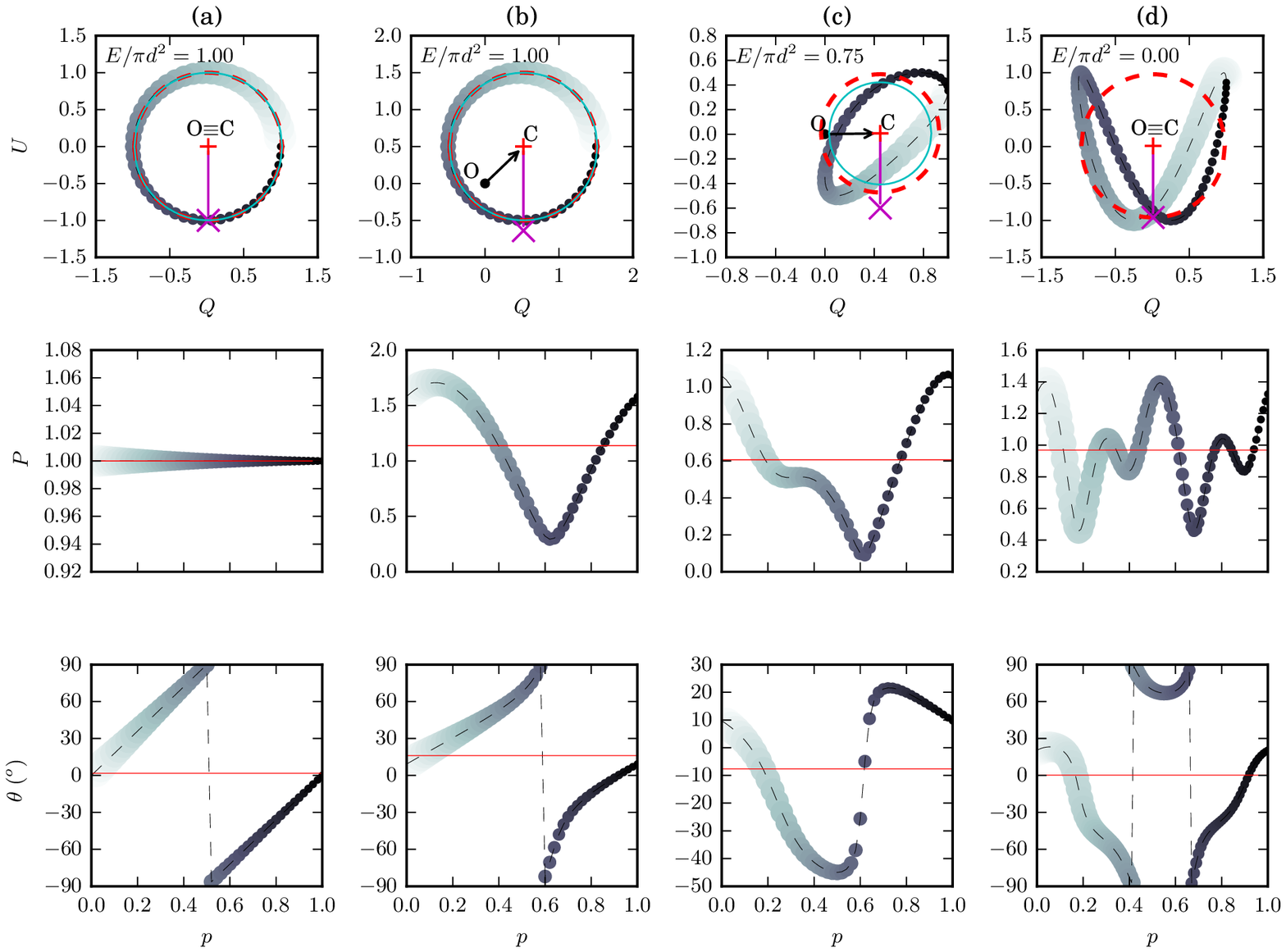}
\caption{Each column shows the $QU$ diagram (top), and the linear polarisation $P$ (middle) and PA (bottom) as functions of time, for closed $\QU$ loops of four different shapes. The size of each $\QU$ point decreases and its shading becomes darker with increasing time. In the $P$ and PA panels the horizontal lines mark their mean values. In the $QU$ diagrams, the origin O of the coordinate system (black circle) and the central point C of the $\QU$ loop (+) are pointed out.
The arrow shows the distance between O and C, while the vertical line from C to the cross sign ($\times$) indicates the mean level of linear polarisation $\mean{P}$. The dashed circle has a radius equal to the mean distance $\mean{d}$ of the $\uqp$ points from C, while the radius of the smaller solid circle equals the mean standard deviation $\sic$ of the distance of the $\uqp$ points from C.
The ratio of the area enclosed by the $\QU$ loop over the area of the dashed circle is annotated on the top of each panel. \newline
(a)~Circle centred at the origin O, so that the distance $P$ of the $(Q,U)$ points from O remains constant. So $P(p)=P_\text{c}=\mean{P}$ and the cross lies on the circle with radius $\mean{d}$. Since $U(p)/Q(p)$ is constant, the PA as a function of phase $p$ has a constant slope, while there are two angle reversals: one abrupt at $p=0.5$, and one where the PA smoothly passes from PA=0 at $p=0$ (or $p=1$).
(b) Circle centred at a point other than the origin, with nearly sinusoidal variations in $P$. There is one abrupt angle reversal at a vertical tangent, but the derivative $\diaf\theta/\diaf p$ is not constant. The mean distance $\mean{d}$ from C is not equal to $\mean{P}$. (c)~Non-circular loop with constant direction of time flow and $\sic$ a little smaller than $\mean{d}$. (d) Loop with one intersection and C outside the loop. The direction of time flow is not constant.\newline
The solid circle with radius equal to $\sic$ coincides with the circle of radius $\mean{d}$ for cases (a) and (b), while it vanishes for (d).}\label{f:deci}
\end{figure*}

In temporal $\QU$ diagrams, if the variation of polarisation is periodic (and does not have any additional trend), the curve connecting all $\uqp$ points of a period will form a closed loop (\fgr\ref{f:deci}, top panels). The mean polarisation level during the cycle is given by
  \begin{equation} \mean{P} = \left<\sqrt{Q^2+U^2}\right> .\end{equation}

Let $\text{C}\equiv(\mean{Q},\mean{U})$ be the central point of the closed curve. The polarisation at C is equal to the distance of C from the origin $\text{O}(0,0)$ of the coordinate system:
  \begin{equation} P_\text{c} =\sqrt{\mean{Q}^2+\mean{U}^2}
                \neq \mean{P}.\end{equation}
The mean distance of the $\uqp$ points from O does not necessarily coincide with the distance of the central point C from O. In the special case that the $\QU$ loop is circular (see \fgr\ref{f:deci}, examples a+b), the point C lies at the centre of the circle. When the $\QU$ loop is circular and $\text{C}\equiv\text{O}$ (\fgr\ref{f:deci}a), which is true for pole-on stars (\fgr\ref{f:quB}), then $P$ is constant ($\mean{P}=P_\text{c}$).
In circular loops, when C does not coincide with O, \mbox{$\mean{P}\neq P_\text{c}$} and $P^2$ varies sinusoidally (\fgr\ref{f:deci}b).

The mean distance $\mean{d}$ of the $\uqp$ points from C is equal to 
  \begin{equation}
  \mean{d} = \left<\sqrt{(Q-\mean{Q})^2+(U-\mean{U})^2}\right>.
  \end{equation}
The variance $\vari(P)$ of linear polarisation $P$ is given by \eqt\eqref{e:stde},
  \begin{equation}
  \vari(P) = \left<|P-\mean{P}|^2\right> \neq \mean{d}^2.
  \end{equation}
When referring to the variance (or standard deviation) in relevance to temporal $\QU$ diagrams, it may be more appropriately defined as the variance $\sic^2$ of a function of two variables ($Q$ and $U$) from the central point C \citep{Papo65}:
  \begin{equation} \vari_\text{c}=\vari(P_\text{c})
  = \left<|P-P_\text{c}|^2\right> =\int_0^1(P-P_\text{c})^2\diaf p.
  \end{equation}
In circular loops the quantity $\sic$ is equal to the mean distance $\mean{d}$ from C to each one of the points $P(Q,U)$. The loop shape diverges from circular as much as the area $E=\upi\sic^2$ enclosed by the circle of radius $\sic$ diverges from $\upi \mean{d}^2$ ($\upi\sic^2<\upi \mean{d}^2$; \fgr\ref{f:deci}). Expanding $\vari(P)$ and approximating $\mean{P}\simeq P_\text{c}$, it can also be shown that \mbox{$\vari(P)<\sic^2$}.

A circle around the central point $(\mean{Q},\mean{U})$ with radius equal to the mean distance $\mean{d}$ from C can serve as a reference in characterising the periodicity of the variation. Plotting the two circles (one with radius $\mean{d}$ and one with radius $\sic$) makes it easy to distinguish the possible shapes of a closed $\QU$ loop.
\begin{enumerate}
\item If the loop maintains its rotational direction (as in panels \ref{f:deci}a, b and c), then $E=\upi\sic^2\leq\upi \mean{d}^2$. In the special case that the loop is circular, $E=\upi \mean{d}^2$ (panels \ref{f:deci}a and b).
\item If the loop does not maintain its rotational direction around C, then circularity is evidently destroyed and $E<\upi\sic^2$. A change in the rotational direction of a $\QU$ diagram implies intersections.
By definition, if a closed curve changes direction around its central point C, the area $E$ of the polygon (i.e.\ closed curve formed by distinct points in an orthogonal coordinate system), is equal to the absolute value of the area enclosed in the clockwise part minus the counter-clockwise part (i.e.\ the so-called Green's theorem, whose proof was given by \citealt{Riem67}).
\fgr\ref{f:deci}(d) is the extreme case where the clockwise part has area exactly equal to the counter-clockwise part, so the algebraical sum of the two is zero ($E=0$).
\end{enumerate}

An ``angle reversal'' or ``angle rotation'' by definition is a change of the PA sign. Steep angle reversals occur when the PA sign abruptly passes from negative to positive values or the contrary. In a plot of PA versus time, steep angle reversals would be seen as vertical segments of the PA curve, i.e.\ at points of infinite derivatives $\partial (\text{PA})/\partial t$ and vertical tangents.
When the vertical tangent lies on the left of the central point C in a $\QU$ diagram, then the change of sign occurs abruptly. When the vertical tangent lies on the right side of C, then the PA changes sign passing smoothly from PA=0. The dinstinction between smooth and abrupt angle reversals might change at a different convention for the PA values (see \fgr\ref{f:geom}b).

The above description points out that $\QU$ diagrams are simple parametric plots of two parameters ($Q$ and $U$), and reveals the connection of their shape and statistics as periodic quantities. The shape of a $\QU$ plot provides information on their periodicity and the inclination $i$ of the line of sight with respect to the stellar axis.
Comparing to \fgrs\ref{f:quB} and \ref{f:quM}, it can be inferred that the $\QU$ diagrams from \fgr\ref{f:deci}(a) to (d) approximately match periodical $\QU$ plots of increasing $i$: The pole-on views of \fgr\ref{f:quB} resemble example (a) centred at the origin O. At a little higher inclinations they become more like (b), as $\mean{Q}$ increases. Then the shape of the $\QU$ loop becomes more elliptical as in example (c).
At even higher inclinations, intersections start and the enclosed area might even become equal to zero as in (d).

\end{document}